\documentclass[reprint,twocolumn,aps,showpacs,amsmath,amssymb,floatfix,superscriptaddress,longbibliography]{revtex4-2}
\usepackage{amsthm}
\usepackage{amsmath, mathtools}
\usepackage[utf8]{inputenc}	
\usepackage[english]{babel}
\usepackage{amsmath}
\usepackage{graphicx}		
\usepackage{natbib}
\usepackage{textcomp}
\usepackage{gensymb}
\usepackage[usenames,dvipsnames,svgnames,table]{xcolor}
\usepackage[hidelinks,colorlinks=false,urlcolor=Cerulean,citecolor=black]{hyperref}
\usepackage{siunitx}
\usepackage{mathrsfs}
\usepackage{multirow}
\usepackage{bm}
\usepackage{xfrac}
\usepackage{comment}
\usepackage[normalem]{ulem}
\usepackage{ragged2e}

\definecolor{yello}{HTML}{f2ffcc}

\newcommand{\C}{\mathbb{C}}

\renewcommand{\i}{\mathrm{i}}

\begin{document}

\title{A mathematical language for linking fine-scale structure in spikes from hundreds to thousands of neurons with behaviour}

\author{Alexandra N. Busch}
\thanks{co-first authors}
\affiliation{Department of Mathematics, Western University, London ON, Canada}
\affiliation{Western Institute for Neuroscience, Western University, London ON, Canada}
\affiliation{Fields Lab for Network Science, Fields Institute, Toronto ON, Canada}

\author{Roberto C. Budzinski}
\thanks{co-first authors}
\affiliation{Department of Mathematics, Western University, London ON, Canada}
\affiliation{Western Institute for Neuroscience, Western University, London ON, Canada}
\affiliation{Fields Lab for Network Science, Fields Institute, Toronto ON, Canada}

\author{Federico W. Pasini}
\affiliation{Huron University College, London ON, Canada}

\author{Ján Mináč}
\affiliation{Department of Mathematics, Western University, London ON, Canada}
\affiliation{Fields Lab for Network Science, Fields Institute, Toronto ON, Canada}

\author{Jonathan A. Michaels}
\affiliation{Department of Physiology and Pharmacology, Western University, London ON, Canada}
\affiliation{School of Kinesiology and Health Science, York University, Toronto ON, Canada}

\author{Megan Roussy}
\affiliation{Department of Physiology and Pharmacology, Western University, London ON, Canada}
\affiliation{Robarts Research Institute, Western University, London ON, Canada}

\author{Roberto A. Gulli}
\affiliation{Columbia University, New York NY, USA}

\author{Benjamin W. Corrigan}
\affiliation{Department of Physiology and Pharmacology, Western University, London ON, Canada}
\affiliation{Robarts Research Institute, Western University, London ON, Canada}

\author{J. Andrew Pruszynski}
\affiliation{Western Institute for Neuroscience, Western University, London ON, Canada}
\affiliation{Department of Physiology and Pharmacology, Western University, London ON, Canada}

\author{Julio Martinez-Trujillo}
\affiliation{Department of Physiology and Pharmacology, Western University, London ON, Canada}
\affiliation{Robarts Research Institute, Western University, London ON, Canada}

\author{Lyle E. Muller}
\email{lmuller2@uwo.ca}
\affiliation{Department of Mathematics, Western University, London ON, Canada}
\affiliation{Western Institute for Neuroscience, Western University, London ON, Canada}
\affiliation{Fields Lab for Network Science, Fields Institute, Toronto ON, Canada}

\begin{abstract}

Recent advances in neural recording technology allow simultaneously recording action potentials from hundreds to thousands of neurons in awake, behaving animals. However, characterizing spike patterns in the resulting data, and linking these patterns to behaviour, remains a challenging task. The lack of a rigorous mathematical language for variable numbers of events (spikes) emitted by multiple agents (neurons) is an important limiting factor. We introduce a new mathematical operation to decompose complex spike patterns into a set of simple, structured elements. This creates a mathematical language that allows comparing spike patterns across trials, detecting sub-patterns, and making links to behaviour via a clear distance measure. We first demonstrate the method using Neuropixel recordings from macaque motor cortex. We then apply the method to dual Utah array recordings from macaque prefrontal cortex, where this technique reveals previously unseen structure that can predict both memory-guided decisions and errors in a virtual-reality working memory task. These results demonstrate that this technique provides a powerful new approach to understand structure in the spike times of neural populations, at a scale that will continue to grow more and more rapidly in upcoming years.

\end{abstract}

\maketitle

\begin{figure*}[tbh]
    \centering
    \includegraphics[width=0.925\textwidth]{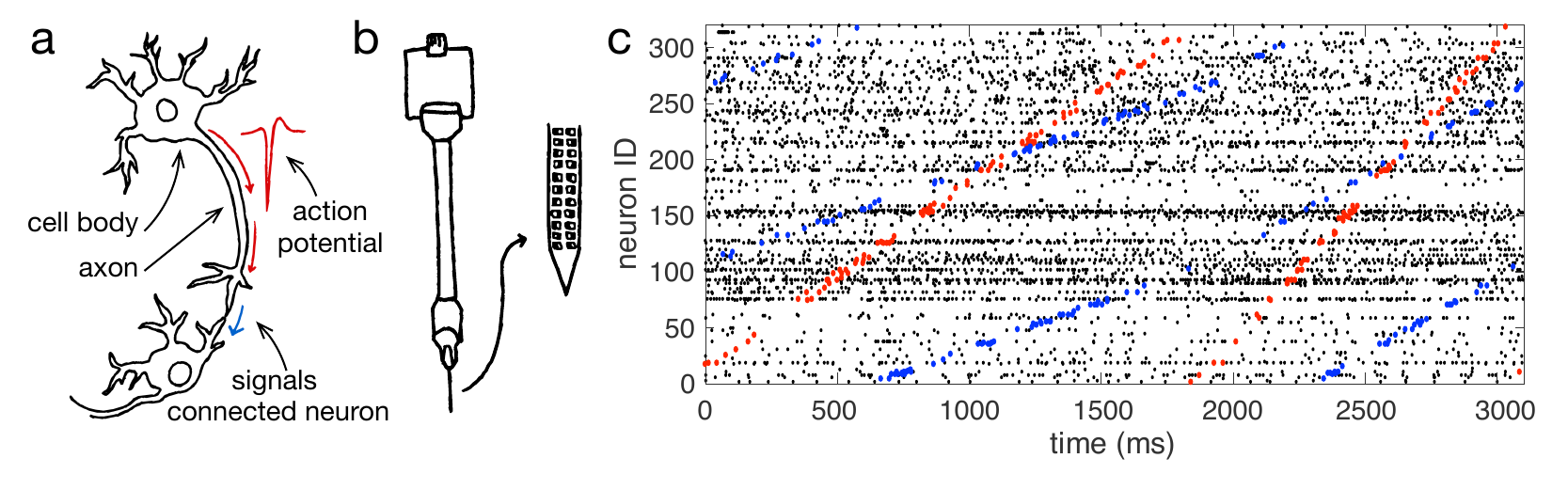}
    \caption{\textbf{Decomposing large-scale spiking data into structured   sub-patterns. (a) } Schematic of connected neurons. An action potential travels down the axon to signal connected neurons. \textbf{(b)} Schematic of a Neuropixel probe. \textbf{(c)} Three seconds of spiking data recorded from a Neuropixel probe implanted in the motor cortex of a macaque monkey performing a reaching task. Several repeating, structured sub-patterns are present in this recording. Spikes contributing to two of these sub-patterns are coloured in red and blue respectively. By extending this analysis, the complete spike pattern can be decomposed into structured sub-patterns by determining which spikes align with basis spike-pattern elements.}
    \label{fig:f1_new}
\end{figure*}

Neurons process sensory inputs, perform computation, and generate motor output via action potentials, or ``spikes''. These discrete events are brief one-milisecond electrical impulses that form the basis for communication between neurons in mammalian brains \cite{bear2020neuroscience}. Researchers often route an electrode recording directly into a speaker, allowing them to find neurons by hearing spikes as distinct ``click" sounds. Neuroscientists have been listening to the ``clicks" of single neurons since the 1920s (Fig.\,\ref{fig:f1_new}a) \cite{adrian1929discharge}. New technologies, however, such as Utah arrays and Neuropixel probes, have rapidly increased the number of neurons that can be recorded simultaneously (Fig.\,\ref{fig:f1_new}b) \cite{jun2017fully, steinmetz2021neuropixels}. Recordings from multiple Neuropixel probes now open the possibility to link patterns in the spikes of hundreds — or even thousands — of neurons to sensory processing, motor behaviour, and cognition \cite{steinmetz2019distributed, durand2023acute}. Despite these advances in recording from many neurons, however, it remains unclear how neurons may coordinate their spikes to support sensory, motor, and cognitive computations. 

Many studies have investigated the temporal evolution of spiking activity across neuronal populations (see for example \cite{riehle2000dynamical,broome2006encoding,rabinovich2008transient,steinemann2024direct, mehrotra2024hyperpolarization}). These works have led to two primary hypotheses: the first posits that a true underlying rate varies meaningfully over time to encode information, and that spikes are stochastic samples of this rate. Many robust methods, such as state-space analysis \cite{srinivasan2006state}, population decoding \cite{quian2009extracting}, and factor analysis \cite{churchland2012neural}, enable testing this hypothesis in large-scale spike recordings \cite{stringer2024analysis}. The second hypothesis is that meaningful coordination may exist in the spikes across a population of neurons at the level of individual trials. Several studies have investigated this possibility \cite{mainen1995reliability, fellous2004discovering, butts2007temporal, kayser2010millisecond, mackevicius2012millisecond, xie2024neuronal}, but it remains difficult to systematically analyze spike patterns in large neural populations, and to link these patterns to behaviour. The lack of a rigorous mathematical language for variable numbers of events (spikes) emitted by multiple agents (neurons) is the limiting factor. This methodological gap means that, while it has been possible to test the first hypothesis in large-scale spike recordings, the second hypothesis remains relatively untested.

To illustrate the difficulty of addressing the second hypothesis, consider a population of hundreds to thousands of neurons. This population will emit many spikes, with these ``clicks'' from individual neurons overlapping in complex patterns over time. If a neuroscientist were to play the ``clicks'' from all these neurons through a speaker, the result would be like listening to a thousand telegraphs sending messages all at once. If the telegraphs encode messages independently, then we have some hope of recovering the encoded messages; however, if messages are interwoven across telegraphs, as would happen if neurons in a population coordinate fine-scale patterns in their spikes, no mathematical method currently exists to detect and quantify structure in this complex series of overlapping events.

A tool for breaking down complex spike patterns into simpler elements would make it possible to detect repeating, structured sub-patterns in neural activity, just as we might identify coordinated messages across telegraphs (Fig.\,\ref{fig:f1_new}c). Such a decomposition could also ask whether fine-scale structure in spike patterns may be related to behaviour and cognition in experimental tasks. The key challenge that has hindered such a method results from the difficulty in adapting the mathematics underlying standard decomposition techniques to discrete spiking data from thousands of neurons. 

Mathematical decomposition techniques rely on computing the ``match'' between real-world input data and a set of simpler elements. In general, this ``match'' is computed by the inner product, a mathematical operation that requires two input vectors to have the same number of entries. However, it is unclear how to compute the inner product between two spike patterns, because neurons fire variable numbers of spikes across trials, resulting in spike patterns of different sizes. To avoid this challenge, it is possible to smooth the spikes from each cell and estimate a continuous firing rate \cite{churchland2012neural, williams2018unsupervised, mackevicius2019unsupervised}, or to limit to a single spike (or burst) from each neuron \cite{gollisch2008rapid, yiling2023robust, xie2024neuronal}. While these approaches allow computing the inner product with two vectors of the same size, they lose the full temporal resolution of the recording, potentially blurring structure in spikes across cells.

Here, we introduce a novel mathematical transformation that solves this fundamental challenge. This transform can be applied to spike patterns from tens of milliseconds to seconds in length, and across thousands of cells, resulting in a mathematically precise description for the fine-scale structure in large-scale spike patterns. This transform provides a natural way to decompose spike patterns into simple, orthogonal elements, which in turn offers a way to detect structured sub-patterns and to rigorously compare spike patterns across trials. Critically, these analyses set no restrictions on the number of spikes or neurons that can be studied, and do not require temporally smoothing the spike data to understand population dynamics.

Computational algorithms have been developed for studying spike patterns in these large data sets. These algorithms tend to focus on detecting coincident spiking events \cite{grun2002unitary, grun2002unitary2, kreuz2013monitoring} and sequences \cite{mackevicius2019unsupervised}, or clustering neurons with similar responses for visualization \cite{vargas2015spike, stringer2024rastermap}, often using distance metrics designed for single neuron spike trains \cite{victor1996nature, van2001novel} or low dimensional embeddings \cite{grossberger2018unsupervised,sotomayor2023spikeship}. However, these methods often require specifying the temporal precision \textit{a priori}, or varying other free parameters. By contrast, the mathematical decomposition we introduce uses a fixed set of simple elements, thus discovering structure in spike patterns with a specific and fixed mathematical operation that is clearly and rigorously interpretable.

By solving the key technical problem with the inner product, the mathematical framework we introduce here provides a novel technique to analyze the large-scale spike recordings that are becoming increasingly important in neuroscience. We demonstrate the utility of this approach by providing example applications to Neuropixel and Utah array data from macaque monkeys. Notably, applying this method to spikes from prefrontal cortex (PFC) allows predicting both choices and errors during a virtual reality (VR) working memory task. Taken together, these initial applications to large-scale spike train recordings demonstrate that this technique provides a powerful new approach to understand structure in the spike times of neural populations, at a scale that will continue to grow more and more rapidly in upcoming years. 

\section{Results}

\subsection{A mathematical language for spike patterns}\label{sec:mathematical_language}

Mathematical transforms can be useful for revealing structure in data that is difficult to observe in one domain, but simple and evident in a different form. Here, we introduce a transform designed to reveal structure in spike patterns from large neural populations. We define a \textit{spike pattern} to be a set of (spike time, neuron ID) pairs that describes the full set of discrete spikes fired across a neural population during some window of time (see Methods Sec. \ref{sec:cylindrical_geometry}). Our goal is to decompose arbitrarily complex spike patterns into simple, structured sub-patterns for further analysis. To achieve such a decomposition, five key features are required:
\begin{itemize}

    \item \textbf{Requirement 1:} The transform must operate directly on the recorded spikes, without smoothing or averaging in any way.
    
    \item \textbf{Requirement 2:} The transform must operate on a set of structured elements that both enables decomposing spike patterns and offers a meaningful interpretation. Satisfying {\bf R2} enables decomposing spike patterns into simple elements that have a clear meaning in terms of spike trains across many neurons.
   
    \item \textbf{Requirement 3:} The transform must be applicable to recordings from hundreds to thousands of neurons, and scale well to next-generation recordings from 10,000 neurons and beyond.

    \item \textbf{Requirement 4:} The transform must have no restriction on the number of spikes per neuron.

    \item \textbf{Requirement 5:} The spike decomposition must exist in a well-defined mathematical space, with a clear distance metric. Satisfying {\bf R5} enables comparing two spike patterns across a large population, for example during two different trials with the same sensory stimulus.

\end{itemize}

The transform we introduce here is the first to satisfy all five of these  requirements. By fulfilling each of these requirements, the transform enables a mathematically rigorous analysis of fine-scale structure in large-scale spiking data, for the first time. We call this operation a {\bf multi-sample Discrete Helix Transform (ms-DHT)}, a name which reflects the key properties of the transform that are described below.

\subsection{Multi-sample Discrete Helix Transform (ms-DHT)}\label{sec:ms_dht}

The goal of a decomposition operator is to break down something complicated into simpler parts. It works by computing the ``match" between each of the simple elements and the original, more complicated, pattern. To define a decomposition for spike patterns, then, it is necessary to decide on the set of simple elements that will power the decomposition, and on how to measure the match. 

\begin{figure*}[htb]
    \centering
    \includegraphics[width=\textwidth]{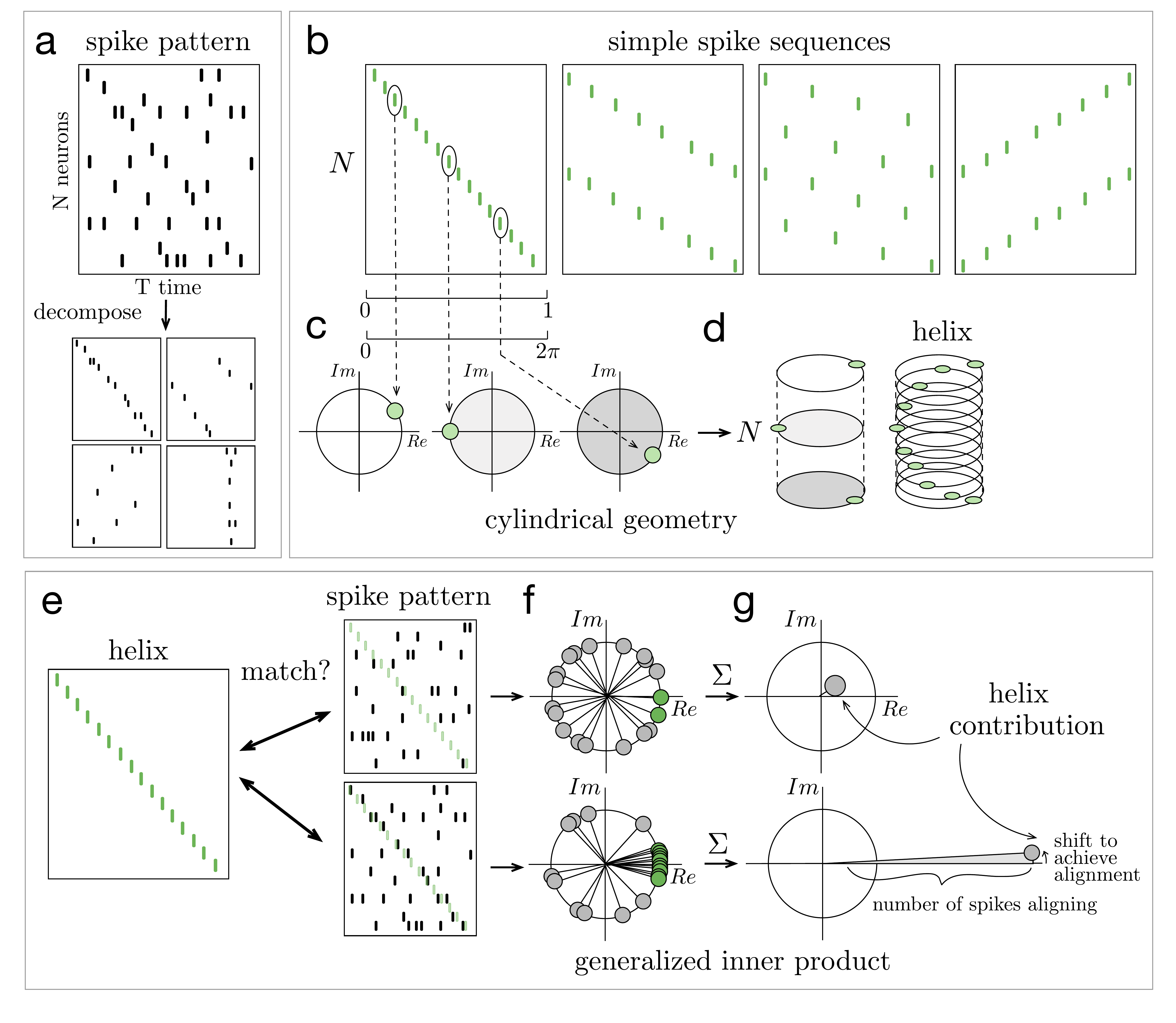}
    \caption{\textbf{Decomposing spike patterns. (a)} Schematic of a spike pattern across $N$ neurons and $T$ seconds being decomposed into structured subpatterns. \textbf{(b)} Our method decomposes spike patterns into a set of simple spike sequences, or ``helices". In the cylindrical geometry that powers this decomposition, spikes are mapped to the unit circle in the complex plane \textbf{(c)} and the circles corresponding to each neuron are stacked vertically into a cylinder \textbf{(d)}. In this geometry, the simple spike sequences become helices wrapping around the cylinder with specific spacing. \textbf{(e)} The decomposition operation measures the match between each helix and the spike pattern. Here, a schematic of this procedure is depicted for two example spike patterns. In the first (top), few spikes align with the helix (green). In the second (bottom), many spikes align closely with the helix (black spikes aligning with green transparent helix). \textbf{(f-g)} Schematic of computing the generalized inner product. \textbf{(f)} Each dot represents the match between one spike in the spike pattern and the helix. Green points (near the positive real axis) correspond to spikes that align closely with the helix. \textbf{(g)} The sum of these points measures the generalized inner product between the spike patterns and the helix, which we term the ``helix contribution".  The magnitude of this sum reflects the strength of the alignment between the helix and the spike pattern (larger magnitude for bottom spike pattern). The phase of this sum reflects the shift required to achieve such alignment.} 
    \label{fig:f2_transform}
\end{figure*}

The ms-DHT decomposes the spike pattern across a neuronal population over a window of time into a set of spike sequences (Fig.\,\ref{fig:f2_transform}a). Sequences are easily interpretable in terms of spikes across a population of neurons. Drawing inspiration from the sinusoidal basis underlying the Fourier transform, the spike sequences we consider here progressively increase in spatial frequency, and traverse the neural population in different directions (Fig.\,\ref{fig:f2_transform}b).  

The underlying geometry of this transform is a cylinder (Fig.\,\ref{fig:f2_transform}c, see Methods Sec.~\ref{sec:cylindrical_geometry}). In this geometry, each simple element or spike sequence is a {\it helix} that wraps around the cylinder with a specific spacing (Fig.\,\ref{fig:f2_transform}d). The vertical axis of the cylinder is made up of integers representing neuron ID, while each circular cross-section is given by the unit circle in the complex plane. The position of each spike around these unit circles is obtained by transforming the time of the spike (within a window of analysis) into a rotation around the cylinder (Fig.\,\ref{fig:f2_transform}c). This geometry then allows defining a method for measuring the match between spike patterns and helices that ensures mismatches are effectively cancelled out, while still detecting matches that are shifted or warped (see Methods Sec.~\ref{sec:gen_ip}, Supplementary Movie 1). This, in turn, makes it possible to detect partial and overlapping matches, to unravel structured sub-patterns that may be interwoven across neurons and time.
\begin{figure*}[bth]
    \centering
    \includegraphics[width=\textwidth]{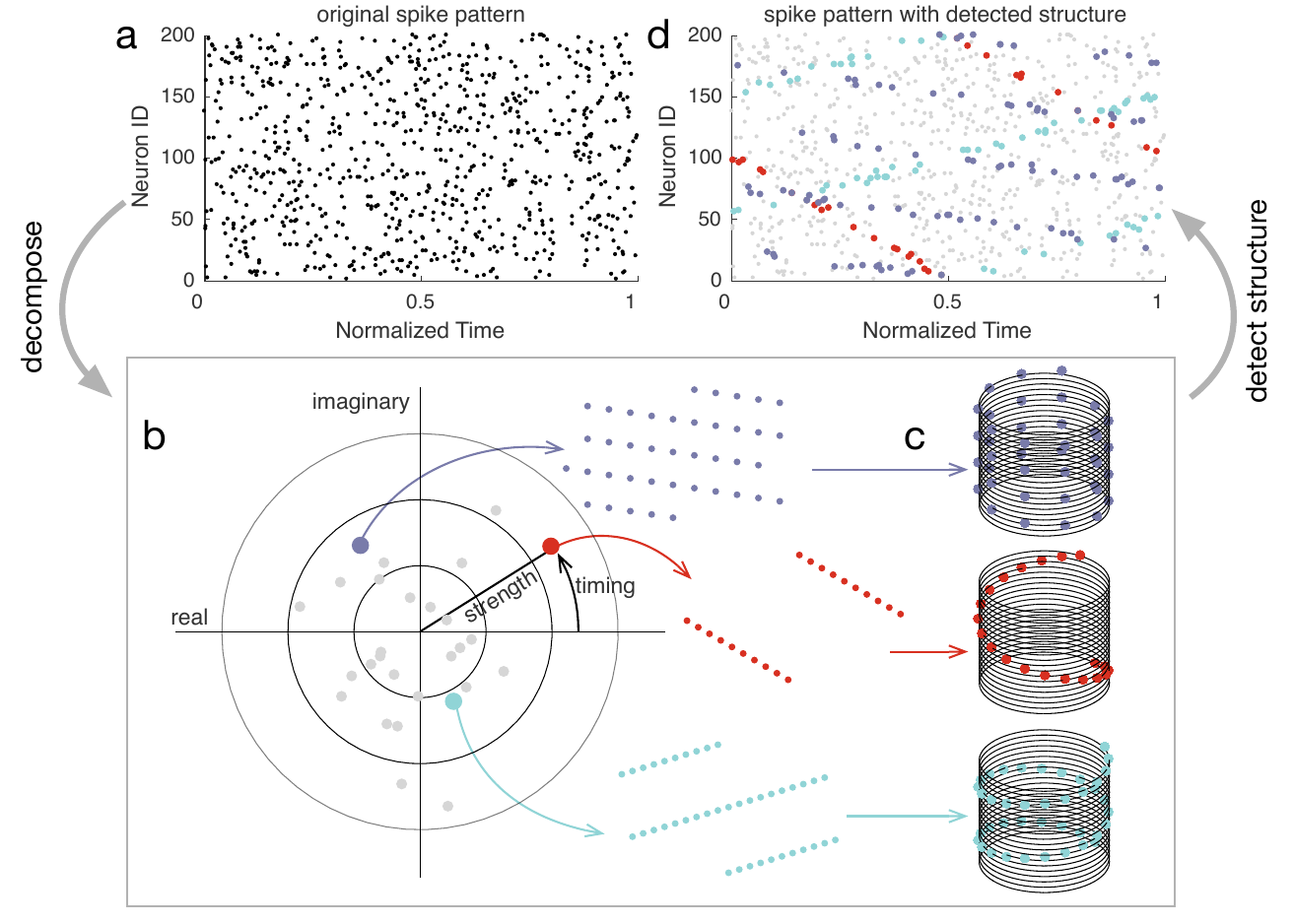}
    \caption{\textbf{Decomposing spike patterns. (a)} Simulated spike pattern across 200 neurons. \textbf{(b)} We begin by measuring the contribution of each basis helix using the generalized inner product (see Methods Sec.~\ref{sec:gen_ip}). The contributions of a subset of helices are plotted in grey. Coloured dots denote the contribution of the example helices plotted in the corresponding colour. For example, the red dot in panel (b) depicts the contribution of the corresponding red basis helix in panel (c). The position of this red point in the complex plane describes the extent to which this red helix is present in the original spike pattern (a). The spikes causing the contribution of this basis helix are coloured red in (d). Similarly, the blue and purple dots reflect the contributions of the blue and purple helices, while the grey dots reflect the contributions of other basis helices. The full collection of points in the complex plane provides a complete description of the spike pattern in (a). \textbf{(c)} Example helices plotted in the cylindrical geometry that powers the ms-DHT. \textbf{(d)} Spikes contributing to the example helices from (b,c) plotted in the corresponding colour code. These sub-patterns can overlap in time and across neurons. }
    \label{fig:f2_new}
\end{figure*}

The process of “breaking down” the original spike pattern works by computing the match between each helix and the whole pattern (Fig.\,\ref{fig:f2_transform}e). Computing this match is precisely where the challenge of decomposing spike patterns arises. Typically, a measurement of similarity, such as the inner product, is used.  However, the inner product depends on comparing objects of the same size, moving element by element to compute a match.  For this reason, previous analyses of population spike patterns have restricted to considering a single spiking event per cell (e.g. first spikes \cite{gollisch2008rapid, yiling2023robust} or spike bursts \cite{yada2016state, xie2024neuronal}), which imposes a fixed size on all spike patterns. As soon as one neuron fires a second spike, however, this operation breaks down: what do you do with that extra spike? Adapting the mathematics of decomposition to the case of spike trains across large neural populations, with no restrictions on the number of spikes fired by any neuron, thus requires developing a new mathematical technique to generalize the idea of the inner product to this case. 

Here, we introduce a generalized form of the inner product to overcome this challenge (see Methods Sec.~\ref{sec:gen_ip}, Eq.~(\ref{eq:mu_k})). The standard inner product moves element-wise along two vectors, always comparing the elements at the same position in each vector. The \textbf{generalized inner product}, on the other hand, moves spike by spike through a spike pattern (Fig.\,\ref{fig:f2_transform}f), and uses the corresponding neuron ID to determine the element of the helix to which each spike should be compared (Supplementary Movie 1). This indexing operation is the key technical detail which allows the generalized inner product to compute the match between a helix and an arbitrary spike pattern of any size. The resulting decomposition measures the extent to which each helix is present in the spike pattern, capturing both the strength of the match (amplitude) and its timing (phase) (Fig.\,\ref{fig:f2_transform}g). 

In summary, the ms-DHT takes a pattern of spikes across many neurons (Fig.\,\ref{fig:f2_new}a), and breaks it down into a set of structured sub-patterns (Fig.\,\ref{fig:f2_new}b), called \textit{helices}, defined in terms of an underlying cylindrical geometry (Fig.\,\ref{fig:f2_new}c). We term the result of this decomposition for a single helix (Fig.\,\ref{fig:f2_new}b, red dot) the \textit{contribution} of that helix to the original spike pattern because it measures the strength and timing of the match between that helix and the original pattern. The vector of contributions across the full set of helices corresponds to the multi-sample Discrete Helix Transform (ms-DHT). This vector (depicted in the complex plane in Fig.\,\ref{fig:f2_new}) provides a description of the fine-scale structure in a spike pattern, which we term its {\it fingerprint}. Note that fingerprints are unique for a given labeling of the neurons (see Methods Sec.\,\ref{sec:mathematical_details}); however, relationships between fingerprints are preserved if we re-label the neurons. This means that an identified link to behavior is not specific to a specific labeling, and that the link will not be impacted by a re-ordering of the cells. A key feature of the ms-DHT, due to the generalized inner product, is that the decomposed spike sequences can overlap in complex patterns across both neurons and time (Fig.\,\ref{fig:f2_new}d).

\subsection{Spike patterns in macaque motor cortex}
\label{sec:motor_cortex}

\begin{figure*}[h!]
    \centering    
    \includegraphics[width=0.8\linewidth]{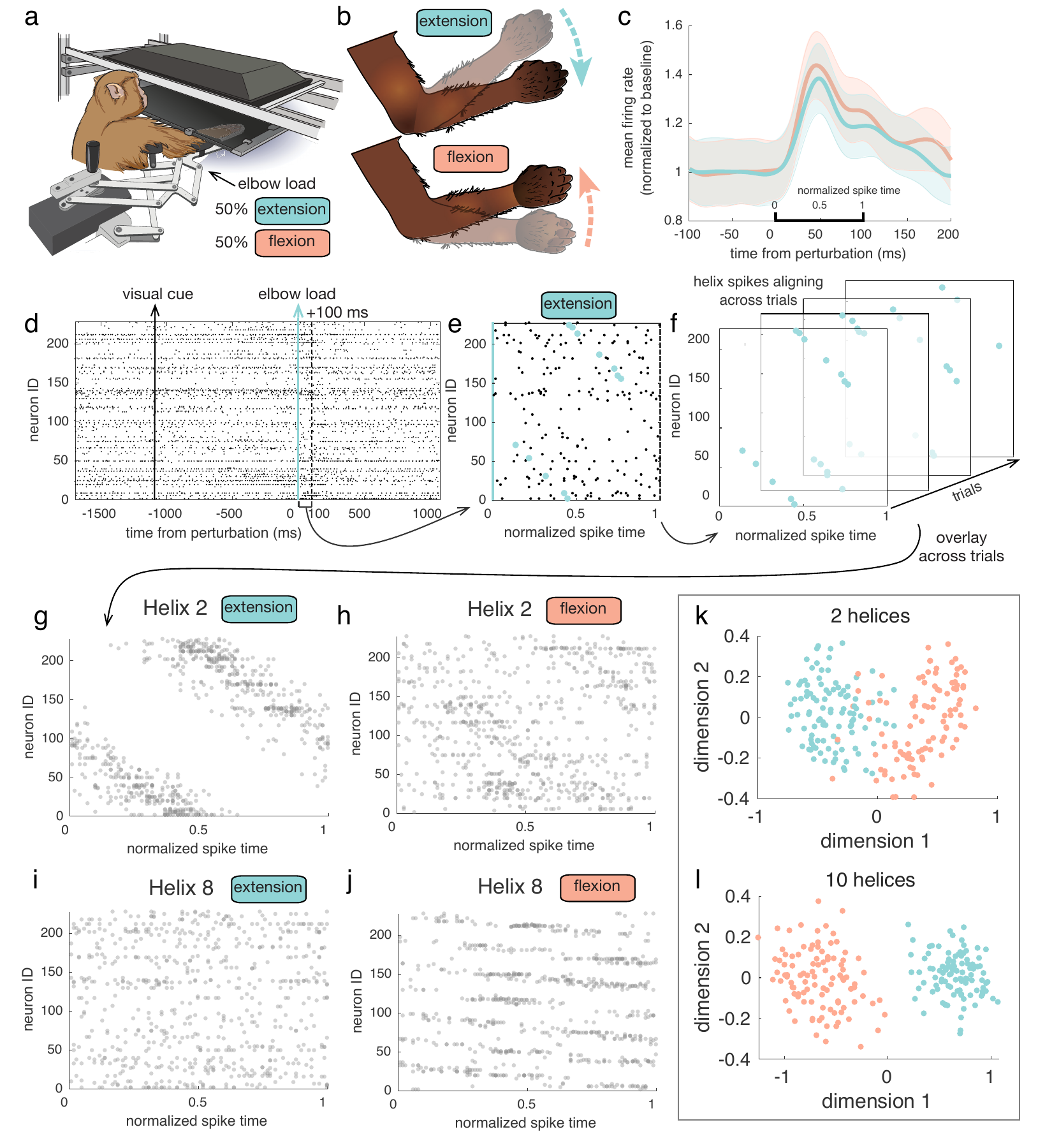}
    \caption{\textbf{Application to Neuropixel data from a motor task. (a)} Task set up. Here, we consider a motor task where a perturbation is applied to the subject's arm by a robotic exoskeleton. The perturbation occurs in one of two directions with equal probability, and at a random time following the start of the trial. \textbf{(b)} Depiction of perturbations. These perturbations are large and very salient, causing either an extension or a flexion of the elbow.  \textbf{(c)} Transient firing rate response. Perturbations are followed by a transient increase in firing rates across the population. Here the mean firing rate across neurons and trials of each perturbation type is plotted, aligned to the time of perturbation, with the shaded region given by the standard deviation. \textbf{(d)} Single trial raster. Following the start of each trial, a visual cue is presented that indicates a perturbation will occur. This cue provides no information about the type of elbow load that will be applied. After a random amount of time, the elbow load is applied (here, an elbow extension, indicated by the blue line). We study the spike pattern in the 100ms window following this perturbation (blue to black dashed line). \textbf{(e)} Single trial spike pattern. Spikes contributing to helix 2 on this trial are highlighted in blue. \textbf{(f)} Four example extension trials are overlaid in the depth dimension. On each trial, the subset of spikes contributing to helix 2 are plotted in blue.  
    \textbf{(g)} Spikes aligning with helix 2 on all extension trials are overlaid with low transparency. Helix 2 was detected reliably on extension trials. Dark regions indicate spikes that repeat reliably across trials. Note that this is not a complete raster: subsets of spikes, corresponding to the blue spikes in panels e and f are plotted for each trial. \textbf{(h)} Spikes aligning with helix 2 on all flexion trials are overlaid with low transparency. Helix 2 was not detected on flexion trials. \textbf{(i)} Spikes aligning with helix 8 across extension trials are overlaid with low transparency. Helix 8 was not detected on extension trials. \textbf{(j)} Spikes aligning with helix 8 on flexion trials. Helix 8 was detected reliably on flexion trials. \textbf{(k)} Clustering based on 2 helices. The perturbation on single trials can be predicted from this projection with 93$\%$ accuracy using a simple SVM. \textbf{(l)} Clustering based on 10 helices. The perturbation on single trials can be predicted from this projection with 99.5$\%$ accuracy using a simple SVM.}
    \label{fig:fneuropixel}
\end{figure*}

As a first demonstration of applying this method to experimental data, we consider Neuropixel recordings from macaque area M1 during a motor task \cite{michaels2024sensory}. The task set up is depicted in Fig.\,\ref{fig:fneuropixel}a. On each trial, a robotic exoskeleton applies a perturbation to the subject's arm in one of two directions, chosen at random with equal probability. These perturbations cause either an extension or a flexion of the elbow (Fig. \,\ref{fig:fneuropixel}b), and occur at a random time following the start of the trial. In the 100ms following these large, salient perturbations, the average firing rate across neurons in this area increases transiently (Fig.\,\ref{fig:fneuropixel}c). Activity in area M1 during this short timescale response is known to encode this type of salient perturbation \cite{omrani2016distributed}, and during this time, the subject instinctively corrects for the elbow load to return their arm to the starting location. In what follows, we decompose and analyze the spike patterns during this $100$ms transient response to the perturbations. This represents a clear and simple demonstration of the approach in a dataset where differences in spiking activity are expected to occur across conditions \cite{omrani2016distributed}. 

On each trial, we study the set of spikes immediately following the perturbation (Fig.\,\ref{fig:fneuropixel}d, blue to dashed black line). We take the ms-DHT of the spike pattern, and study the resulting helix contribution. We found that some helices were detected more frequently on trials of a specific perturbation direction (i.e.~$||\mu_k||$, for a specific $k$, was high relative to chance level for one condition). For example, in this session, helix 2 (Fig.\,\ref{fig:fneuropixel}d-h) was detected more often following extension perturbations (blue), while helix 8 (Fig.\,\ref{fig:fneuropixel}i,j) was detected more often on flexion trials (pink). To visualize this structure, we can find the set of spikes on each trial that contribute to this helix being detected (Fig.\,\ref{fig:fneuropixel}e, blue spikes), in the same manner as before, in simulated data (see for example Fig.\,\ref{fig:f2_new}, red spikes and red helix). Across extension trials when helix 2 was detected with high magnitude, the helix also occurred with similar timing (Fig.\,\ref{fig:fneuropixel}f). In other words, on extension trials when $||\mu_2||$ was large, $\text{Arg}(\mu_2)$ was similar. This repeated structure can be visualized by overlaying the spikes across extension trials that align with helix 2 (Fig.\,\ref{fig:fneuropixel}f,g). When these spikes are overlaid across all extension perturbation trials with low transparency (Fig.\,\ref{fig:fneuropixel}g), dark regions indicate spikes that repeated reliably across trials, with some variability in timing (indicated by ``smearing" in horizontal direction). On the other hand, when the spikes contributing to helix 2 across flexion trials are overlaid, no such structure is visible (Fig.\,\ref{fig:fneuropixel}h). On these trials, $||\mu_2||$ is small, and $\text{Arg}(\mu_2)$ varies greatly; however, a different helix was reliably detected. Helix 8 was detected on flexion trials in this session, and occurred with reliable timing across trials. As before, this helix is readily apparent in the spikes across flexion trials (Fig.\,\ref{fig:fneuropixel}j), but is not visible in the spikes of extension trials (Fig.\,\ref{fig:fneuropixel}i). These results demonstrate how the ms-DHT can be used to detect robustly repeating and behaviourally relevant substructure in spike patterns, where the magnitude and phase of a helix contribution represents the amount and timing of such structure, respectively (as depicted in Fig. \,\ref{fig:f2_new}b). 

These two example helices demonstrate perturbation-specific structure in spike patterns that can be detected using the ms-DHT. These two helices occur reliably enough across trials that they can be used to predict the perturbation on an individual trial with 93$\%$ accuracy (Fig.\,\ref{fig:fneuropixel}k). By including additional helices, the spike patterns group into even more separable clusters, increasing decoding accuracy to 99.5$\%$ (Fig.\,\ref{fig:fneuropixel}l). Interestingly, this perturbation-specific structure can be found immediately following the perturbation in area M1, but not in PFC (see Supplement Fig.~S2). Taken together, these results demonstrate that output of the transform can be used to reliably detect patterns that repeat robustly across trials (as in Fig.\,\ref{fig:fneuropixel}g,j). These patterns can then be traced back to the spikes on single trials (as in Fig.\,\ref{fig:fneuropixel}e,f). Further, these results also demonstrate that the transform can reliably distinguish cases where structure is present (as in Fig.\,\ref{fig:fneuropixel}g,j, and Supplement Fig. S3 area M1) from cases where no such structure exists (as in Fig.\,\ref{fig:fneuropixel}h,i, and Supplement Fig. S3 PFC).

\subsection{Spike patterns in macaque prefrontal cortex}
\label{sec:pfc}

\begin{figure*}[htb]
    \centering
    \includegraphics[width=\textwidth]{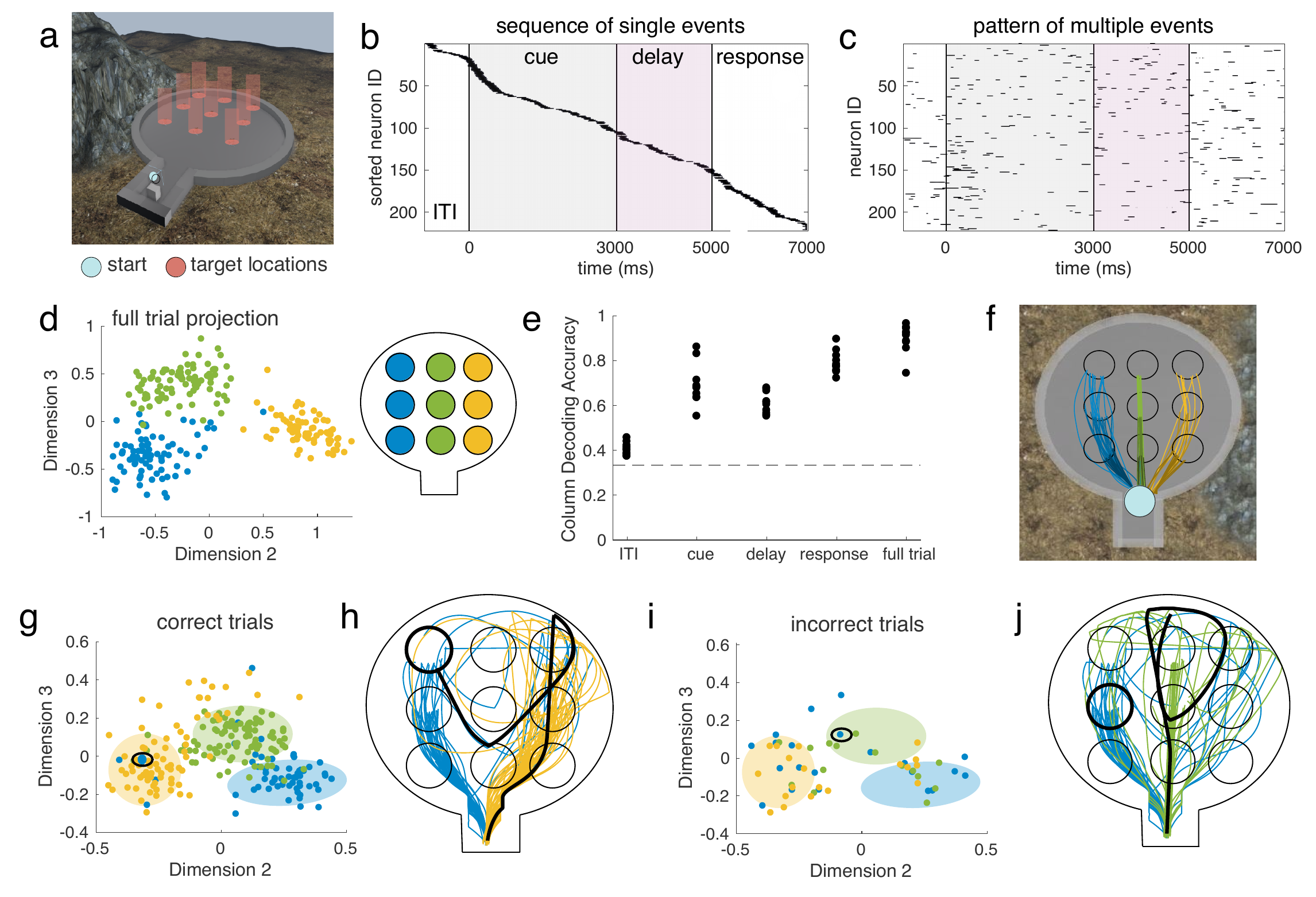}
    \caption{{\textbf{Spike patterns in macaque lPFC on correct trials of a working memory task. (a)} Virtual environment for working memory task. \textbf{(b)} In previous work, we demonstrated that sequences of neuronal activations encoded WM content during this task. \textbf{(c)} Here, we can extend our previous analyses to include multiple firing events per cell. \textbf{(d)} Full trial spike patterns on correct trials cluster by column. colour code depicted in a schematic of the virtual environment from above. \textbf{(e)} Column can be decoded from the position of single trials within the projection with high accuracy throughout the task, except for during the ITI when the target has not yet appeared on screen. \textbf{(f)} In previous work, we noticed that subjects tended to take similar paths to targets within the same column, which could explain this clustering. Example correct trajectories to each target, plotted on an overhead view of the virtual environment. \textbf{(g)} Projection of correct trial spike patterns for one session during navigation. Circled: an example blue (left column) trial positioned in the yellow (right column) cluster. \textbf{(h)} The path taken by the subject on this example trial is plotted in black, alongside other correct paths to blue and yellow targets. \textbf{(i)} Projection of incorrect trials coloured by column for the same example session. Circled: an example blue (left column) trial positioned in the correct green (center column) cluster (green circle, green dots in panel g). \textbf{(j)}  The path taken on this example trial is plotted in black, alongside example blue and green trial paths.}}
    \label{fig:f5}
\end{figure*}

Next, we applied this method to recordings from two Utah arrays implanted in the left lateral prefrontal cortex (lPFC) of a macaque monkey performing a working memory (WM) task that takes place in a virtual environment (Fig.\,\ref{fig:f5}a) \cite{roussy2022stable}. In the task, one of nine target locations is cued by a visual stimulus, which then disappears for a two second delay period. After the delay, the subject navigates to the remembered location using a joystick. The nine targets are arranged into a grid of three rows (each consisting of three targets from left to right at a specific depth) and three columns (left, right, and center). 

The recordings from two Utah arrays yield between 200-300 well-isolated single units in each session. In previous work, we found that sequences of single neuron activations encoded working memory content in this task (Fig.\,\ref{fig:f5}b) \cite{busch2024neuronal}. These sequences were composed of brief 100-200 ms elevations of firing rate well above baseline, and the position of each cell in the sequence was determined by the time it reached its maximum firing rate. The sequences had similar ordering and timing on trials of the same condition, and were separable between conditions. While we previously restricted to consider only one spiking event per cell on each trial, the ms-DHT now allows us to consider the complex pattern of all such elevations across all cells within a trial (Fig.\,\ref{fig:f5}c, Methods Sec. \,\ref{sec:utah_array}). This allows us to determine (1) whether we can replicate our previous results with this method and (2) whether including additional spiking events provides any additional insight.

To address the first question, we confirmed that the ms-DHT recovers the structure across trials found previously. In particular, we previously found that sequences corresponding to targets in the same column clustered together, so that the target column could be easily decoded from the sequences on individual trials. Here, we find that the ms-DHT recovers this clustering by column (Fig.\,\ref{fig:f5}d) and achieves comparable decoding performance to the analysis in the restricted setting (Fig.\,\ref{fig:f5}e). While subjects were navigating to a target, they tended to take similar trajectories for targets within the same column (Fig.\,\ref{fig:f5}f). By expanding our analysis with the ms-DHT, we can now link the complex pattern of spiking events (Fig.\,\ref{fig:f5}c) to the specific trajectory taken on individual trials.

We used the ms-DHT to obtain a fingerprint of the spiking events on each trial, and then used the well-defined distance between fingerprints to study their similarity across trials (see Methods Sec.~\ref{sec:comparing}). Spike patterns during the navigation epoch of correct trials form three distinct clusters based on column (Fig.\,\ref{fig:f5}g), showing that trials where navigation was toward the left column (encoded in blue) exhibited similar spike patterns. Not all trials fell within the cluster for the corresponding column, however: in some trials with a target in the left column, for example, the spike pattern was classified as more similar to those with a target in the right column (Fig.\,\ref{fig:f5}g, blue circled dot in yellow cluster). These ``misplaced'' trials correspond to those with targets in the left column, but with spike patterns that are more similar to trials with targets in the right column. 

To what do these ``misplaced'' trials correspond? When studying the navigation paths subjects took on these ``misplaced'' trials, the navigation paths are at first more similar to the trials with which the spike pattern clustered. Taking for example an individual trial with a target in the left column (encoded in blue), the spike pattern during navigation was more similar to trials with a target in the right column (Fig.\,\ref{fig:f5}g, blue dot in yellow cluster, circled in black). On that trial, the subject initially navigated to the right (Fig.\,\ref{fig:f5}h, black path compared to blue and yellow paths), reached the end of the arena without receiving a reward, then turned around and headed to the left to correct the mistake, eventually reaching the correct target (bold circle). These results extend to incorrect trials, where the similarity between these incorrect trials and the spike patterns on correct trials could explain the specific type of error made on individual trials (Fig.\,\ref{fig:f5}i,j). Finally, these results were consistent across trials, where we found that paths from misclassified trials were significantly more similar to paths in the predicted cluster than to paths in their true cluster (STATS) (see Supplement Fig.~S3). These results show that spike patterns occurring during navigation could more closely reflect the navigation strategy than the end of the trajectory at the correct target (Fig.\,\ref{fig:f5}h,j, bold circle), and that analysis of spike patterns by the ms-DHT can predict the specific errors made on individual trials (see Methods Sec.~\ref{sec:utah_array} for details). 

Taken together, these results demonstrate that the ms-DHT can be flexibly applied across a variety of contexts and timescales. Results from this method are consistent with our previous findings relating simplfied sequential activity patterns to WM content. Finally, the ms-DHT offers additional insights into the choices and errors on individual trials, demonstrating that the method finds behaviorally relevant patterns in complex neural activity~--~in this case, the actual navigation path taken by a subject on individual trials in a VR environment.

\section*{Discussion}
 
In this work, we have introduced a novel framework for characterizing patterns of spiking activity across large populations of neurons. This approach represents the first complete mathematical language for spike patterns, which provides a simple way to describe, compare, and decompose patterns of spikes on a variety of timescales. We provide example applications to spiking data from macaque motor and prefrontal cortex recorded using Neuropixel probes and Utah arrays. These applications demonstrate how the multi-sample Discrete Helix Transform (ms-DHT) can be used to find behaviourally relevant structure in spike patterns that can then be directly related to the spikes on a single trial level. 

\subsection*{A mathematical language for spike patterns}

The ms-DHT satisfies the five requirements we set out to achieve (Results - A mathematical language for spike patterns). \textbf{R1:} \textit{The transform must operate directly on the recorded spikes, without smoothing or averaging.} The ms-DHT operates on sets of discrete spikes, at the level of individual trials. \textbf{R2:} \textit{The transform must operate on a basis that enables decomposing spike patterns and offers a meaningful interpretation}. The cylindrical geometry underlying the ms-DHT allows decomposing spike patterns into simple elements, or \textit{helices}, which are readily interpretable as specific sequences of spikes across a population of neurons. \textbf{R3:} \textit{The transform must scale well to next-generation recordings from 10,000 neurons and beyond.} There is no restriction on the number of neurons which can be included in the cylindrical geometry. Further, since the ms-DHT relies solely on indexing operations and element-wise multiplications, it scales easily to recordings from very large populations. \textbf{R4:} \textit{The transform must have no restriction on the number of spikes per neuron.} The generalized inner product for comparing arbitrary spike patterns to helices represents the key technical detail making it possible to remove all restrictions on the number of spiking events which can be considered. \textbf{R5:} \textit{The spike decomposition must exist in a well-defined mathematical space, with a clear distance metric.} The output of the ms-DHT is a complex-valued vector. Clear distance metrics already exist in this space, which can then be used to compare spike patterns. The example applications presented in this work suggest that measuring the distance between spike patterns in this way can reveal structure that is relevant to both motor and working memory behaviour.  

\subsection*{Difference from previous work}

There has been long-standing interest in methods for analyzing spike trains. The Victor-Purpura distance \cite{victor1996nature} is an early example of a mathematically precise way to characterize the similarity between spike trains of single neurons. This method has provided insight into the temporal resolution of neural coding in sensory areas \cite{victor1997sensory}, but focuses on pairwise comparisons between two spike trains. Scaling to large populations has required reducing the complexity of single neuron spike trains down to one spike time per cell \cite{gollisch2008rapid,yiling2023robust, xie2024neuronal}, utilizing optimization algorithms \cite{sotomayor2023spikeship}, or parametrizing the temporal resolution of the neural code \cite{grun2002unitary,grun2002unitary2, aronov2003neural}. Several algorithms have been designed to visualize hidden structure in large-scale spiking data. For example, algorithms such as Rastermap \cite{stringer2024rastermap}, SIMNETS \cite{hynes2018simnets}, and SSIMs \cite{vargas2015spike} can be used to group neurons with similar spiking responses. 

A variety of methods has been developed to handle large-scale data sets from many simultaneously recorded neurons \cite{stringer2024analysis}. Many of these methods require first smoothing the data, such as neural state-space analysis \cite{churchland2012neural}, and rate-based decoding methods \cite{pouget2000information}. While these types of methods are useful for analyzing the activity of large populations over time, they cannot account for the precise timing of individual spikes, and are therefore unable to investigate the possibility of fine-scale structure in spike timing that could be relevant for supporting cognition and behaviour. These methods have thus been able to test the possibility that spikes are stochastic samples of an underlying firing rate that evolves meaningfully over time. However, there has been a lack of methods available to test the alternate hypothesis, that meaningful coordination could exist in the spikes across a population of neurons.

\subsection*{Decomposition operation for discrete spike patterns}

The cylindrical geometry underlying the ms-DHT leverages symmetries in the complex plane to detect shifted and warped copies of patterns, without the need for computationally expensive time-warping algorithms \cite{williams2020discovering} or combinatorial methods that include cell relabeling as an additional parameter \cite{aronov2003neural}. This allows the method to scale to analyze very large populations, without the computational issues associated with combinatorial methods \cite{grun2002unitary, grun2002unitary2} or those requiring parameter optimization \cite{williams2018unsupervised, vargas2015spike}). Further, this representation allows for the possibility that the timing of spikes relative to the activity of other cells in the population could be important for neural coding. Future work could explore different geometries that may enable different insights. 

The key technical advance presented in this work is a method for decomposing spike patterns, using a generalized form of the inner product that compares complex patterns of many spikes across many cells to simple, structured elements of fixed sizes. Other geometries could be explored in future work, such as a non-periodic geometry that could be more useful for certain types of data. Here, we find that the ms-DHT can be flexibly applied in a variety of cases and provides unique insight into the possibility of previously undiscovered structure in the spiking activity of large populations of neurons. A key feature of the ms-DHT is its interpretability, as it can link behaviourally relevant coordination in spiking activity directly to the spikes on individual trials. Thus, the ms-DHT thus opens new possibilities for flexible analyses of large-scale spike recordings during complex experimental paradigms.   

\subsection*{Future directions}

Our initial applications of the ms-DHT demonstrate the potential for revealing structured patterns in the spikes of hundreds of neurons. These results extend from the motor cortex, where a specific spike pattern corresponds to each of two perturbations, to lPFC, where spike patterns allow predicting both correct choices and errors. Future applications of the ms-DHT could reveal even more sophisticated structure in large-scale spike recordings. Specifically, distinct but overlapping sequences of spiking activity might encode multiple streams of information simultaneously - in the case of the VR WM task above, for instance, items held in working memory and the outcome of previous trials. These overlapping patterns could allow neurons to efficiently represent multiple aspects of behavior at once, providing the computational flexibility and compositionality required for making behaviorally relevant decisions in complex environments. The results presented here demonstrate that the ms-DHT opens new possibilities for understanding compositional, multiplexed codes embedded in the spikes of large neural populations, providing a fundamentally new technique for understanding how neurons coordinate their activity to process sensory information, make decisions, and generate motor output.

\renewcommand\thefigure{M\arabic{figure}}
\setcounter{figure}{0}    

\section{Methods}

\subsection{Cylindrical geometry for spike patterns}
\label{sec:cylindrical_geometry}

\begin{figure}[h!]
    \centering
    \includegraphics[width=\linewidth]{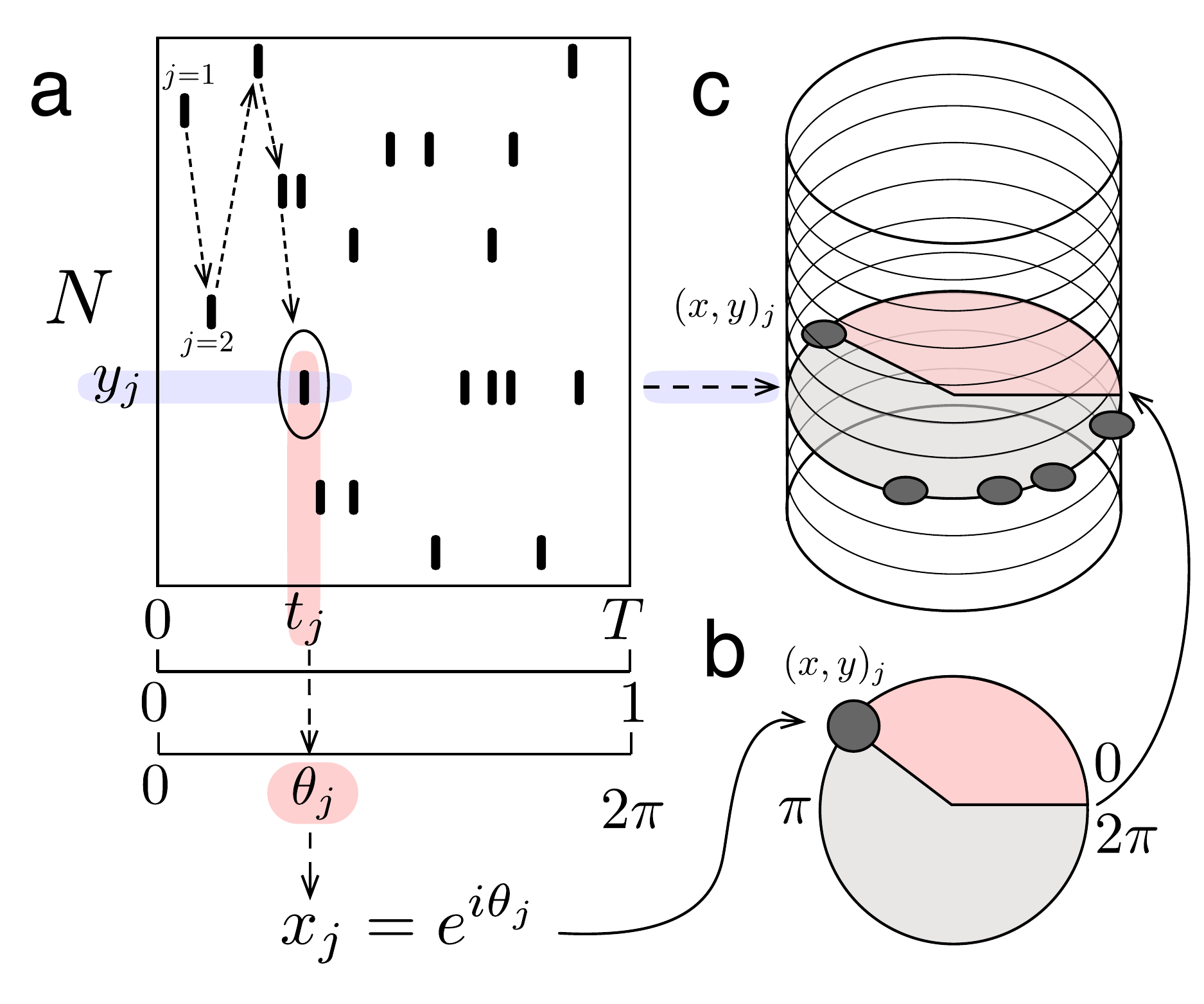}
    \caption{\textbf{Cylindrical geometry (a)} Schematic of example spike pattern across $N$ neurons over $T$ seconds. Spikes are indexed by $j$ in temporal order (indicated by dashed arrows). Transformation from recorded spike time to complex-valued spike demonstrated for example spike $j=5$ (circled). \textbf{(b)} Example spike represented on the unit circle in the complex plane. Red shaded region indicates time from start of window. \textbf{(c)} Example spike depicted on cylinder, along with other spikes from the same neuron (shaded blue arrow, shaded grey circle). }
    \label{fig:Methods1}
\end{figure}

Consider a pattern of $M$ spikes fired by a population of $N$ neurons over some amount of time, $T$ seconds. Each spike in the pattern can be uniquely described by the time of the spike and the ID of the corresponding neuron. In other words, the $j^{th}$ spike in the pattern can be described by the pair $(t,y)_j$, where $t_j \in [0,T]$ is the spike time, $y_j \in [1, N]$ is the neuron ID, and $j \in [1,M]$ indexes spikes in the pattern according to the temporal order in which they occur (Fig.\,\ref{fig:Methods1}a). 

The cylindrical geometry of the ms-DHT uses complex numbers to represent the timing of spikes, by transforming each spike time $t_j$ to the unit circle in the complex plane. To achieve this transformation, time within the window of analysis (from $0$ to $T$) is first normalized (from $0$ to $1$) and then linearly mapped to phase values (from $0$ to $2\pi$ radians), which correspond to points on the unit circle in the complex plane. More formally, each spike time $t_j$ is transformed to the complex valued spike $x_j := e^{\i \theta_j}$, where $\theta_j := 2\pi \i \frac{t_j}{T}$ (Fig.\,\ref{fig:Methods1}b). We note that $\i = \sqrt{-1}$ is the imaginary unit, and $i$ is used as an index in this work.

In this geometry, then, a \emph{spike pattern} is the set $s := \{ (x,y)_j \}_{j=1}^M$ . The $j^{th}$ element $(x,y)_j$ describes the spike $x_j := e^{2\pi i \frac{t_j}{T}}$ fired by the neuron $y_j \in [1,N]$ at time $t_j$, as described above. The spike pattern $s$ is a subset of has a natural representation as a collection of points on a cylinder of radius $1$ and height $N$, where points on the unit circle at each horizontal plane represent the spikes of one neuron and are positioned around the circle according to their timing (Fig.\,\ref{fig:Methods1}c). More formally, $s \subset S^1 \times \{1,...,N\}$, and we define $\mathcal{S}$ to be the set of all such spike patterns.  

\subsection{Decomposing spike patterns}
\label{sec:decomposing}

\begin{figure}[t!]
    \centering
    \includegraphics[width=\linewidth]{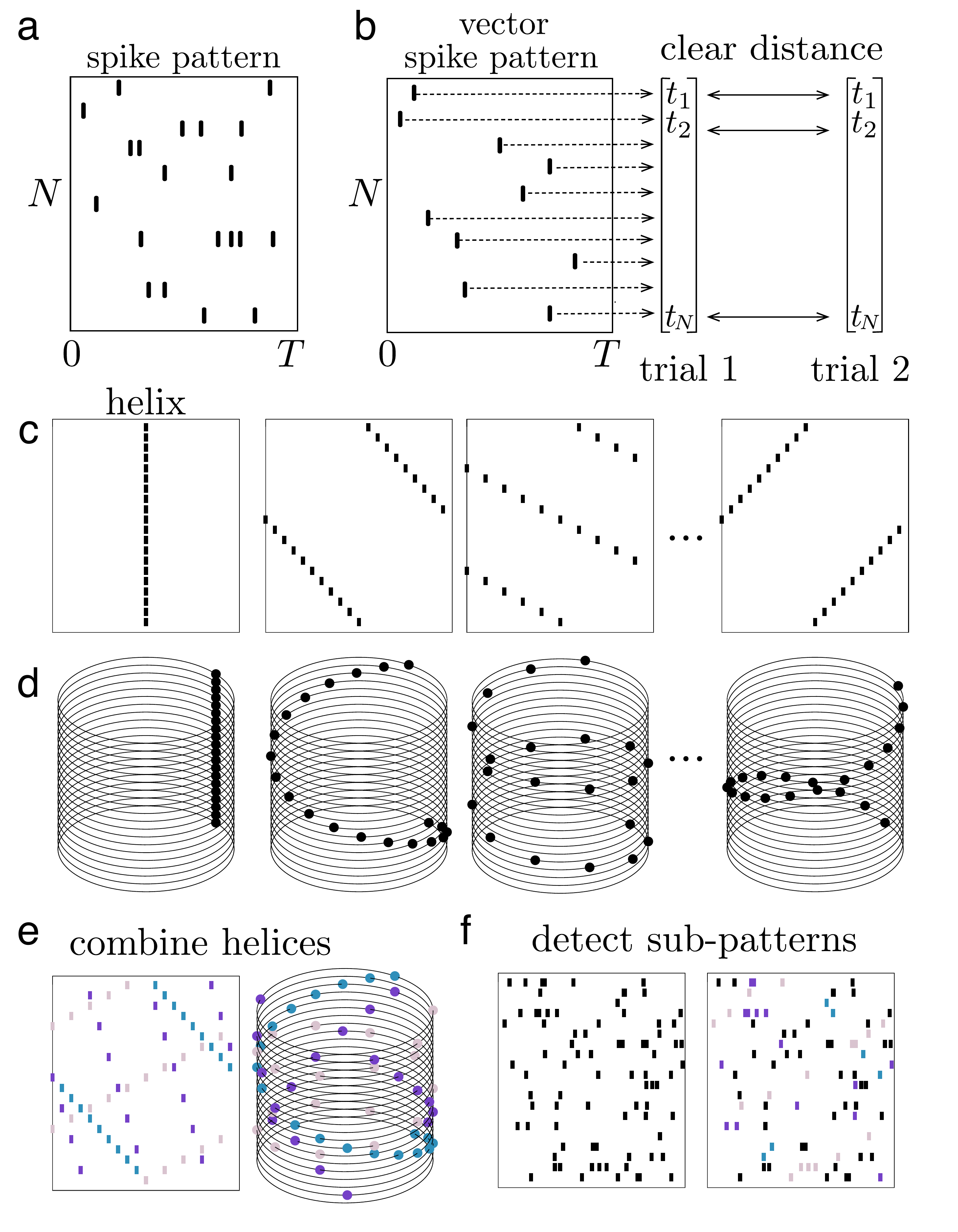}
    \caption{\textbf{Spike patterns and helices. (a)} Example spike pattern across $N$ neurons over $T$ seconds, where neurons fire multiple spikes. \textbf{(b)} Example vector spike pattern, where each neuron fires one spike. Clear distance metrics exist for comparing vector spike patterns because they have fixed sizes. \textbf{(c)} Four example helices, which are vector spike patterns with specific structure, and can be thought of as building blocks for general spike patterns. \textbf{(d)} The same for helices plotted in cylindrical coordinates. \textbf{(e)} Three helices (in colour code) are combined to construct a more complex spike pattern, visualized as a raster plot and on the cylinder. \textbf{(f)} A complex spike pattern with many spikes (black spikes). The helices from (e) can be detected as structured sub-patterns within a more complex spike pattern (spikes highlighted in the same colour code as panel e).}
    \label{fig:Methods2}
\end{figure}

The goal of a decomposition operation is to break down something complicated into simple elements. Here, the goal is to decompose complicated patterns of many spikes across a population of neurons (Fig.\,\ref{fig:Methods2}a). In this context, spike patterns where each neuron in the population fires one spike suggest themselves as a natural choice for simple elements. We refer to these as \textit{vector spike patterns}, because they can be written as vectors of length $N$, where each component of the vector describes the spike of the corresponding neuron (Fig.\,\ref{fig:Methods2}b). Vector spike patterns are easy to compare because there are clear choices for distance measures to compare vectors of the same size. For this reason, many previous methods for studying population spike patterns have restricted to the case of vector spike patterns \cite{gollisch2008rapid,yiling2023robust}. To circumvent this restriction, the ms-DHT decomposes arbitrary spike patterns of any size into a special set of vector spike patterns of fixed size using a generalized version of the inner product designed to accommodate this case (see Methods Sec.~\ref{sec:gen_ip}). 

The special set of vector spike patterns that power the decomposition can be thought of as ``building blocks" of spike patterns across $N$ cells. These building blocks are simple sequences of spikes across the population with specific slopes, ordering, and directions \ref{fig:Methods2}c). In the cylindrical geometry, these sequences correspond to \emph{helices} that wrap either clockwise or counterclockwise around the cylinder with a different spatial frequency (Fig.\,\ref{fig:Methods2}d), and form a discrete Fourier basis, $H$, of size $N$. More precisely, the elements of $H$ are the vectors, or helices, $\vec{h}_k = [\omega^k, \omega^{2k}, ... \omega^{Nk} ] $ where the $j^{th}$ component, $(\vec{h}_k)_j$, defines the spike of cell $j$, and $\omega^k$ denotes the $k^{th}$ $N$-root of unity, $e^{\frac{-2\pi \i k}{N}}$. In this way, the $k^{th}$ helix $\vec{h}_k$ corresponds to the spike pattern $s = \{(x,y)_j : x_j = (\vec{h}_k)_j, y_j = j \}$. Each helix forms a spike pattern across the population with one spike per cell, and they can be combined to construct more complex patterns (Fig.\,\ref{fig:Methods2}e).

We use these helices to decompose complex spike patterns into simple sequences by measuring how well the spike pattern aligns with each helix (see Methods Sec.~\ref{sec:gen_ip}). Decomposing spike patterns in this way provides a language for describing and comparing complex spike patterns across trials. This process can be used to detect individual helices that may appear as sub-patterns within a larger spike pattern (Fig.\,\ref{fig:Methods2}f). Importantly, partial and noisy basis helices can also be detected (see Supplement Fig.~S1). 

\subsection{Generalized inner product}
\label{sec:gen_ip}

Given an arbitrary spike pattern, it is possible to determine whether a given helix is present, and to what extent. The \emph{contribution} of a helix $\vec{h}_k$ to a given spike pattern $s$ is measured using a generalized version of the inner product, given by:
\begin{equation}
    \mu_k = \frac{1}{N}\sum_{j=1}^M x_j \cdot (\vec{h}_k)_{y_j}^* 
    = \frac{1}{N}\sum_{j=1}^M x_j \cdot e^{y_j (\frac{2\pi \i k}{N})}.
    \label{eq:mu_k}
\end{equation}
Here, $h_k$ is a vector of length $N$, $s$ is a pattern of $M$ spikes across $N$ neurons, and $(\vec{h_k})^*$ denotes the conjugate transpose of $(\vec{h_k})$. The generalized inner product uses the neuron ID $y_j$ of each spike in $s$ to index the element of $h_k$ to which that spike should be compared. In this way, it enables a mathematically rigorous comparison between objects of different sizes.  
 
\begin{figure}[htb]
    \centering
    \includegraphics[width=\linewidth]{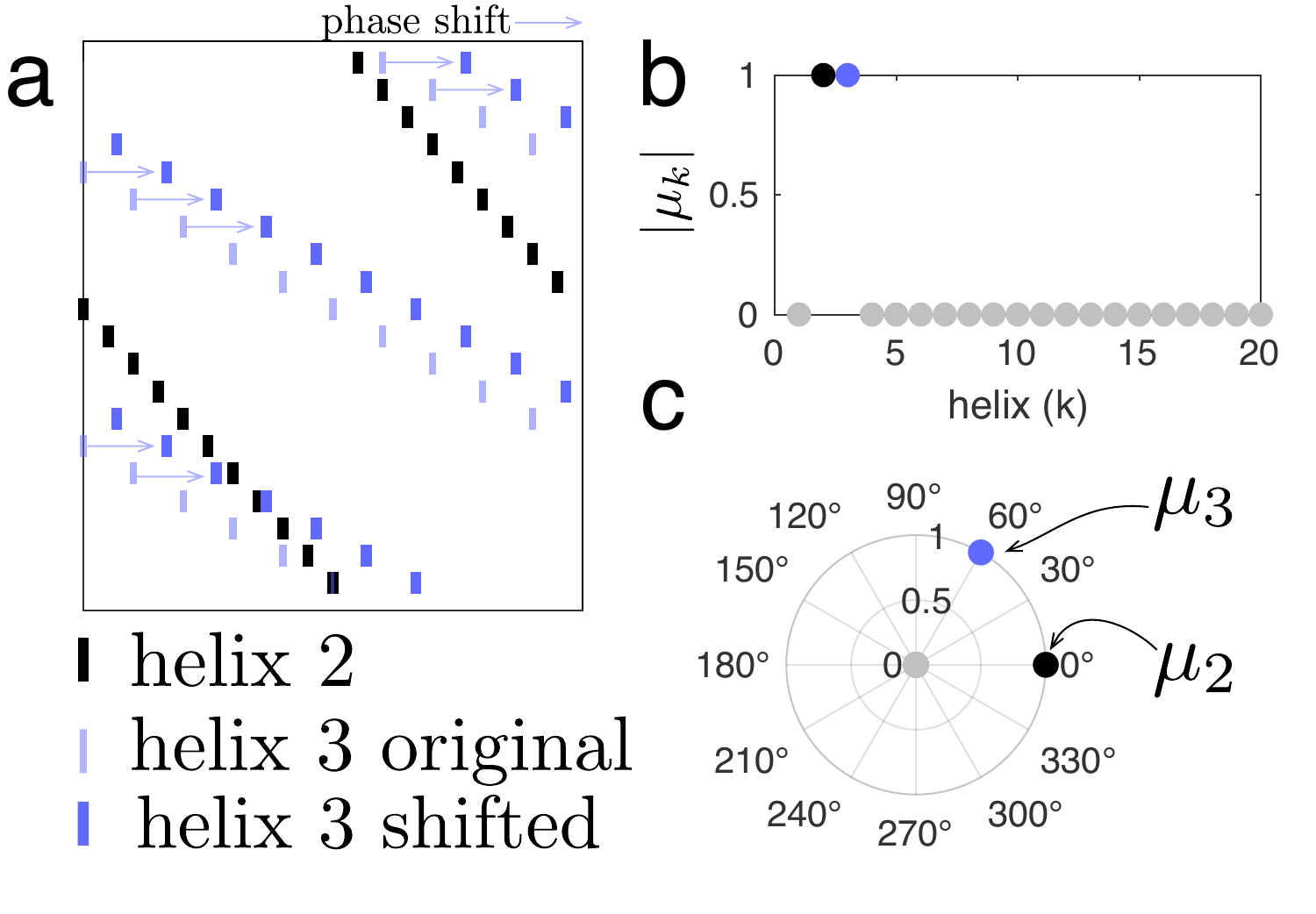}
    \caption{\textbf{Generalized inner product. (a)} Example spike pattern created from helix 2 (black) and a shifted copy of helix 3 (blue, thick lines shifted from original thin lines). \textbf{(b)} Magnitude of helix contributions. Helices 2 and 3 are both detected (magnitude = 1). \textbf{(c)} Polar plot of helix contributions, with radius given by magnitude and angle given by the argument, indicating time shifts.}
    \label{fig:Methods3}
\end{figure}

Note that $\mu_k \in \C$ describes the alignment between helix $\vec{h}_k$ and spike pattern $s$, where $\|\mu_k \|$ describes the fraction of spikes in  $\vec{h}_k$ that align with spikes in $s$, and Arg$(\mu_k)$ describes the phase shift in time required to achieve such alignment. In this way, $\mu_k=1$ precisely when one (and only one) perfect copy of the $k^{th}$ helix is present in a given spike pattern. For example, Fig.\,\ref{fig:Methods3}a depicts a spike pattern created by combining helix $2$ (black spikes) with a shifted copy of helix $3$ (purple spikes). The faint purple spikes represent the original, non-shifted helix $3$ and are plotted just for reference. In that case, $|\mu_2|$ and $|\mu_3|$ equal $1$, while all other helix contributions are $0$; i.e., helices $2$ and $3$ are detected within the spike pattern, but no other helices are detected (Fig.\,\ref{fig:Methods3}b). Further, Arg$(\mu_2)$ and Arg$(\mu_3)$ describe the shifts of helix $2$ and $3$ respectively (Fig.\,\ref{fig:Methods3}c)
 
\subsection{multi-sample Discrete Helix Transform}
\label{sec:msDHT}

To extend the idea of helix contributions to the complete set of helices, we introduce an operator, $\Psi_H : S \to \C^N $, that quantifies the contribution of each helix to a given spike pattern:
\begin{equation}
    \Psi_H(s) := \vec{ \mu },    
      \label{eq:generalized_ip}
\end{equation}
where $ \vec{ \mu }_k := \mu_k$ as defined in Eq. \eqref{eq:mu_k}. In this way, the $k^{th}$ element of $\vec{ \mu }$ describes the contribution of the helix $\vec{h}_k$ to the spike pattern $s$ (Fig.\,\ref{fig:Methods4}). This operator reliably determines the presence of both partial and noisy helices (see Supplement Fig.~S1). Further, by estimating the contribution of each helix to random spike patterns with matched numbers of spikes and neurons, this method can be used to determine whether observed spikes have more specific temporal structure than expected by chance (Fig.\,\ref{fig:Methods4}c, red dots). Finally, this method can be used to decompose arbitrary spike patterns: by measuring the contribution of each helix to a spike pattern, we can obtain a ``fingerprint" describing the temporal structure of that pattern (Fig.\,\ref{fig:Methods4}c). 

\begin{figure}[htb]
    \centering
    \includegraphics[width=\linewidth]{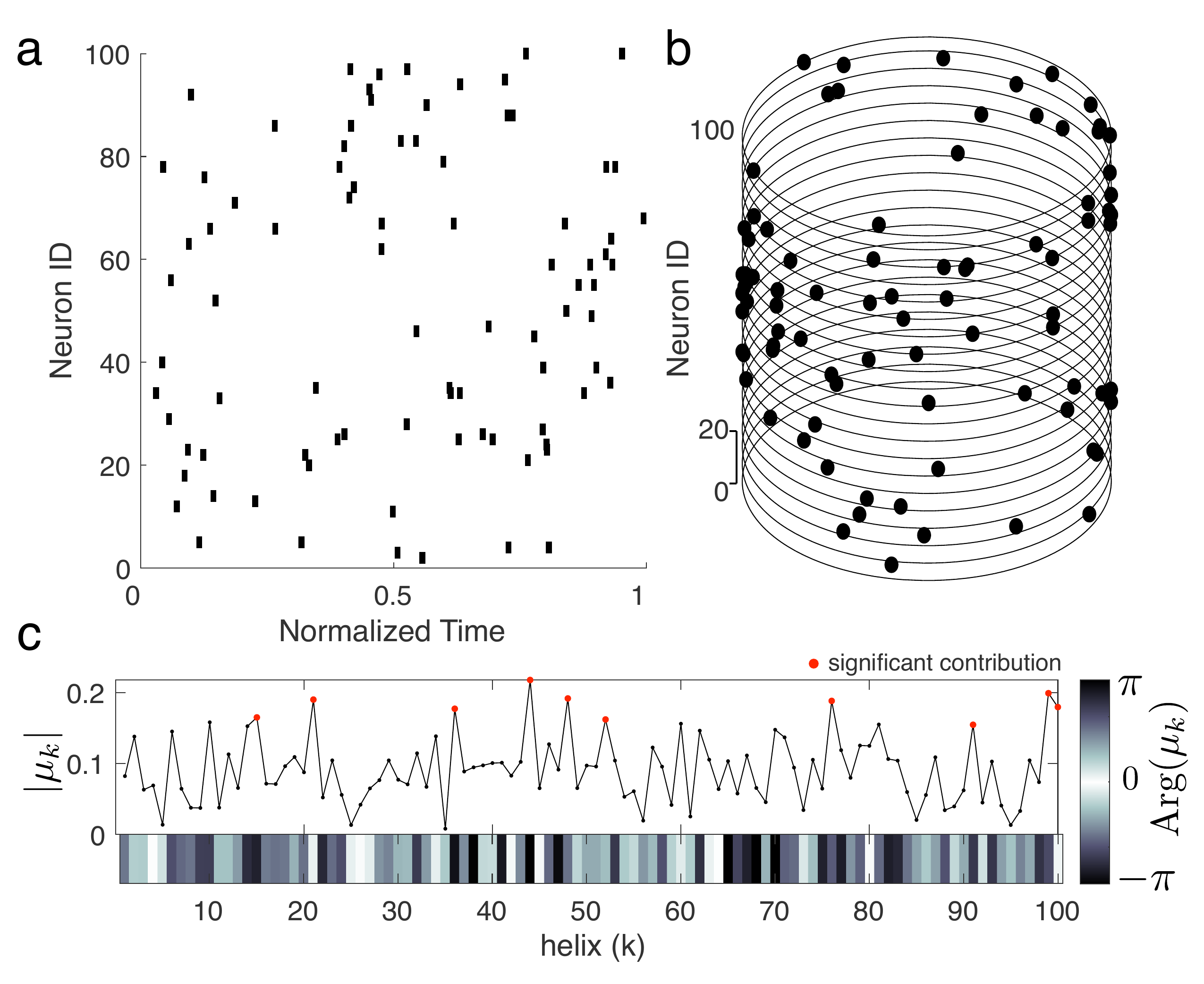}
    \caption{\textbf{ms-DHT.} Example spike pattern across 100 neurons depicted \textbf{(a)} as a raster plot and \textbf{(b)} in cylindrical coordinates. Magnitude of helix contributions. Helices 2 and 3 are both detected (magnitude = 1). \textbf{(c)} The output of the ms-DHT for this spike pattern is a complex-valued vector of length 100, plotted here in terms of magnitude and phase (argument). Red dots denote helices with larger contribution than 1000 random spike patterns with matched numbers of spikes and neurons. }
    \label{fig:Methods4}
\end{figure}

\subsection{Comparing spike patterns}
\label{sec:comparing}

The decomposition operator $\Psi$ reduces an arbitrarily complex spike pattern to a single vector of helix contributions. With this simple representation, it is straightforward to compare spike patterns across trials by comparing the contributions of basis elements to each pattern. For example, one could ask whether trials of a certain condition in an experiment display the same helices with high contributions, or whether dominant helices change systematically with task structure, or perhaps with learning during a task. One could also look at the ``fingerprint" of the spike pattern overall, and group trials with similar spike patterns together to see if certain types of patterns could be related to behaviour. 

With the decomposition provided by $\Psi$, defining a measure of distance between spike patterns is now as simple as selecting a measure of distance in the complex plane to compare the $\vec{\mu}$ vectors across trials. Here, we use the Euclidean distance. It is then straightforward to compute the distances between each pair of spike patterns, and perform dimensionality reduction to discover clusters of similar spike patterns in the data. Here, we use spectral clustering, but other methods are equally applicable.

\begin{figure}[htb]
    \centering
    \includegraphics[width=\linewidth]{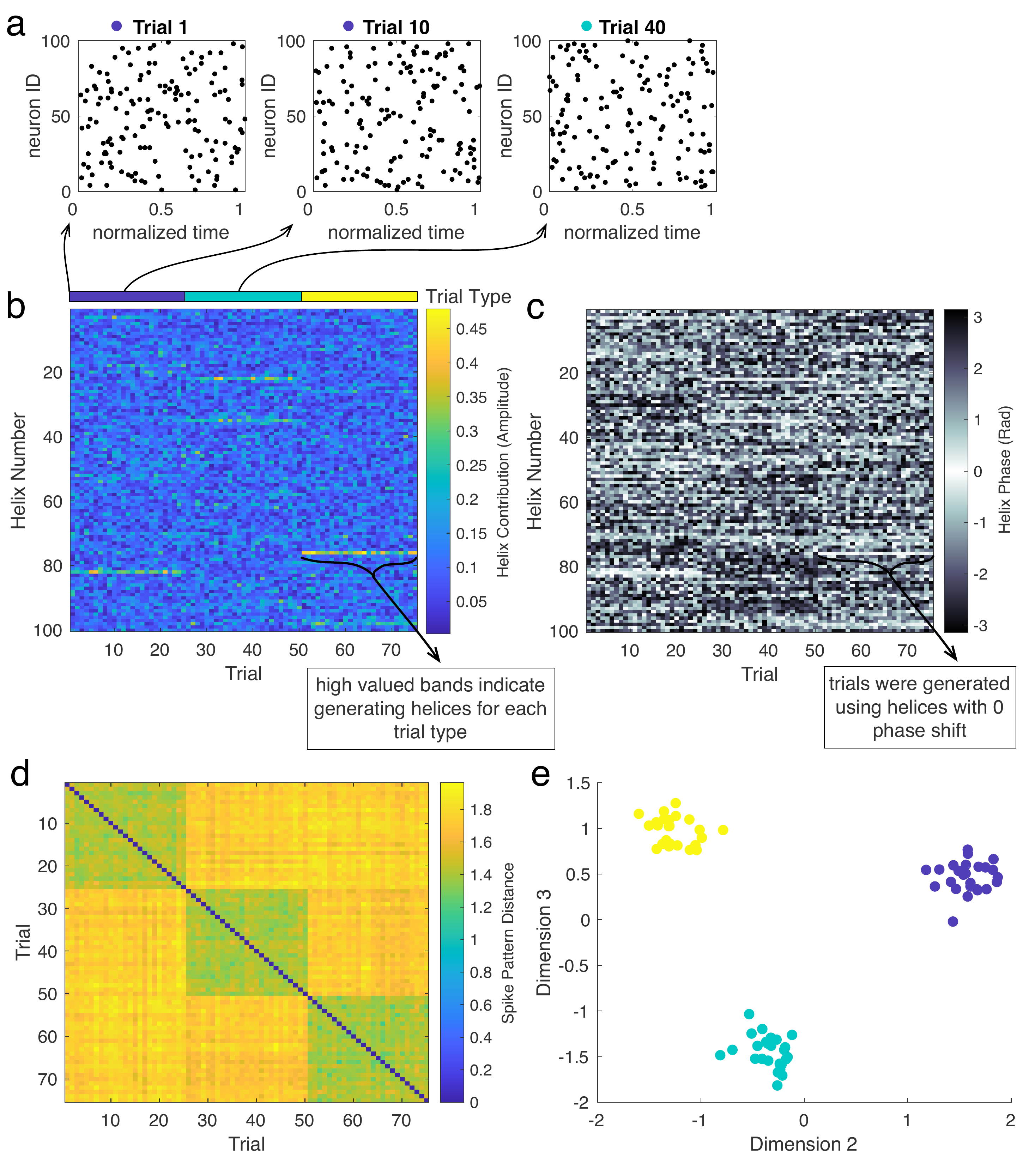}
    \caption{\textbf{Clustering spike patterns.} In this example, there are three different conditions, each one related to the presence of a given helix in the spike pattern. We then consider 75 trials, where each one is generated by a random spike pattern embedded with a partial version of one of the helices. \textbf{(a)} Example trials generated by embedding partial helices into patterns of random spikes. \textbf{(b)} Magnitude of helix contributions across trials, sorted by generating helix (``trial type", also clear in high-valued helix contributions).  \textbf{(c)} Phase of helix contributions. \textbf{(d)} Matrix of distances between spike pattern ``fingerprints" ($\vec{\mu}$) for each pair of trials. \textbf{(e)} Spectral clustering clearly groups trials generated using the same helix. }
    \label{fig:Methods5}
\end{figure}

To demonstrate this point, we first embed partial helices into patterns of randomly generated spikes. We use three different helices to generate $75$ trials of $3$ conditions, where each trial corresponds to a random spike pattern with a partial copy of one of the three helices. We then decompose the spike patterns and perform dimensionality reduction to cluster the trials by condition. Fig.\,\ref{fig:Methods5}a depicts an example of three such spike patterns, where trials 1 and 10 are generated using the same helix, and trial 40 is generated with a different helix. We then use $\Psi$ to decompose the spike patterns on each trial. The amplitude and phase of the resulting complex-valued vectors are depicted in Fig.\,\ref{fig:Methods5}b and c respectively. We then compute the distance between these vectors for each pair of trials. The resulting distance matrix (Fig.\,\ref{fig:Methods5}d) displays three low-valued blocks along the diagonal, corresponding to the three groups of trials generated using the same helices. We then project the similarity matrix onto its own eigenvectors with largest modulus eigenvalues (Fig.\,\ref{fig:Methods5}e, see Sec. \ref{sec:dim_reduction} for more details). In the resulting ``similarity space", each point corresponds to one spike pattern, and distances between points are determined by the distance between spike patterns. This is demonstrated in Fig.\,\ref{fig:Methods5}e, where points are coloured by the helix used to generate the corresponding spike patterns. Using this method, spike patterns clearly cluster based on the embedded helix (i.e., points of the same colour group together). Importantly, this clustering is preserved even when cells are relabeled (see Supplement, Figure S4). 

\subsection{Dimensionality reduction procedure}
\label{sec:dim_reduction}

In more detail, to compare $N_T$ spike patterns across $N$ cells, we begin by computing $\Psi_H(s_i)$ for the spike pattern $s$ on each trial $i \in [1,N_T]$. We next compute the distance between each pair of spike patterns to produce the distance matrix $D \in \mathcal{M}_{N_T \times N_T}$ with $d_{i,j}:= ||\Psi_H(s_i) - \Psi_H(s_j)|| $. We then compute the eigenvalues $\lambda_1, ... \lambda_{N_T}$ and corresponding eigenvectors $\vec{v}_1, ....\vec{v}_{N_T}$, with eigenvalues ordered such that $ |\lambda_1| \geq |\lambda_2| \geq ... \geq \lambda_{N_T}$. Finally, we obtain the  the $n$-dimensional projection $P := D[\vec{v}_1,...\vec{v}_n]$. This projection underlies the decoding in Fig. \ref{fig:fneuropixel}g,h and Fig. \ref{fig:f5}e, as well as the clustering in Fig. \ref{fig:f5}d,g,i. 

\subsection{Cluster distance ratio}
\label{sec:distance_ratio}

Consider a projection with distinct clusters as in (Fig.\,\ref{fig:Methods5}e). The inter cluster distance is defined to be the average distance between cluster centroids. The intra cluster distance is the average distance between points within a cluster to the centroid of that cluster. The distance ratio is then the inter cluster distance divided by the intra cluster distance, where large values correspond to highly separable clusters, and values below one correspond to clusters that are somewhat overlapping.

\subsection{Significantly different helices}
\label{sec:bootstrap_centroids}

To determine whether helix contributions differed across trial conditions (as in Fig.\,\ref{fig:f5}n), we first computed a set of $1000$ bootstrapped centroids for each condition. The helix was considered to differ between conditions if these sets of centroids were non-overlapping (see for example Supplementary Figure 2b). 

\subsection{Application to Neuropixel data}
\label{sec:neuropixel_details}

Experimental details can be found at \cite{Michaels2024bioarxiv}. All procedures described below were approved by the Institutional Animal Care and Use Committee at Western University (Protocol 2022-028). 

Subjects were positioned with their right arm in a KINARM robot exoskeleton (BKIN Technologies \cite{scott1999apparatus}), allowing flexion and extension movement of the shoulder and elbow joints in the horizontal plane. The robot can independently apply specific flexion or torques at these joints. Throughout the experiment, a constant background load of 0.02 Nm extension torque was applied at the elbow. On each trial, monkeys waited with their fingertip in a central target. After a variable delay (600-800 ms), a visual cue was presented that indicated a perturbation would occur. This cue gave no information about the perturbation direction. If, at any point before the perturbation, the hand went outside the home target, the trial was aborted. After a variable delay of 800-1200 ms, monkeys received one of two unpredictable elbow perturbations (±0.2 Nm step-torque) which served as a go cue to compensate for the perturbation and return to the central target. After returning to the central target and holding the hand there for 700 ms, a juice reward was given. 10$\%$ of trials were catch trials where no perturbation was applied. These trials were not included in the spike pattern analysis. 

High-density Neuropixels probe recordings were performed (1.0 - 1 cm, 1.0 NHP - 1 cm, and 1.0 NHP - 4.5 cm). Neural recording targets were identified by registering the CT to a pre-surgery MRI, and identifying the 3D location of each brain area by warping segmentations from a composite macaque atlas to the individual MRI of each animal (NMT v2 \cite{jung2021comprehensive,seidlitz2018population}). Neural data were recorded from Neuropixels probes using SpikeGLX. Neural data were processed using a custom processing pipeline \cite{jonathan_github}. For Monkey M, AP stream data was first drift corrected using spike localization and decentralized registration \cite{boussard2021three,varol2021decentralized} implemented in spikeinterface \cite{buccino2020spikeinterface}, which was able to accurately track vertical probe drift and correct it. Due to the large craniotomy, some of these recordings had large drift (0-250 um). Neural data were then processed with Kilosort 2.0 \cite{pachitariu2024spike} to further stabilize recordings during spike sorting. For Monkey P, drift was minimal due to small craniotomies (drift 0-15 um), so the data was immediately processed using Kilosort 4.0 \cite{pachitariu2024spike}, including built-in drift correction. Single neurons were considered successfully recorded if they were flagged by Kilosort as single neurons using default parameters, and if they were stably recorded for the duration of the recording. 

Data used in this application consisted of $100$ms of spiking activity immediately following the perturbation on each trial. Helices 2 and 8 were selected for visualization because the amplitude of their contributions reliably exceeded chance-level on only one type of perturbation. While a number of other helices also exhibited such selectivity (Supplemental Figure 2), selecting helices with lower spatial frequency aids visualization. 

\textbf{Decoding:} The perturbation on single trials was decoded from the contributions of 2 and 10 of the most reliably occuring, perturbation-selective helices using a support vector machine (SVM). The SVM was trained on vectors of the real and imaginary parts (concatenated) of these helix contributions. Accuracy was reported with 10-fold cross-validation. 

\subsection{Application to Utah array data}
\label{sec:utah_array}

In the task, one of nine target locations is cued with a visual stimulus, which remains on screen for three seconds. The cue period is followed by a two second delay period, where the stimulus disappears. During this time, eye position is unconstrained, but navigation through the environment is disabled. At the end of the delay period, navigation turns on, and the animal uses a joystick to navigate through the environment to the remembered target location to receive a juice reward. 

Experimental details and information about neuronal activation sequences can be found in \cite{busch2024neuronal}. Animal care and handling (i.e., basic care, animal training, surgical procedures, and experimental injections) were pre-approved by the University of Western Ontario Animal Care Committee. This approval ensures that federal (Canadian Council on Animal Care), provincial (Ontario Animals Research Act), regulatory bodies (e.g. CIHR/NSERC), and other national standards (CALAM) for the ethical use of animals are followed.

To extend these sequences to more complicated spike patterns, spike times were defined to be the center of a spiking event when a neuron exceeded $80\%$ of its peak firing rate. Epochs were divided by task timing: ITI = $1$ second before the appearance of the cue; ``cue" = $3$ seconds following cue onset, while the visual cue is present on screen; ``delay" = $2$ second memory delay; ``response" = first $2$ seconds of navigation.  

\textbf{Decoding:} The condition on single trials was decoded from the position of each trial within the projection. The data was divided into 10 folds (training and test sets) for K-Fold cross-validation. For each fold, the training set was used to define the centroid for each condition (the average of points corresponding to trials of that condition). Each trial in the test set was then labelled according to the closest centroid. 

\textbf{Choices and errors:} To quantify the results from Figure \ref{fig:f5}g-j across all trials in ten recording sessions from this subject, we used K-Means clustering to group trials by column. Since K-Means clustering is an unsupervised method based on distances within the projection, and distances within this projection are determined by the distance between spike patterns, we can use the K-Means algorithm to label each trial according to which type of spike pattern that trial resembles most closely. 
As expected, the algorithm clustered some trials correctly (e.g. Fig \ref{fig:f5}k blue dot in blue cluster) and other trials incorrectly (e.g. Fig \ref{fig:f5}k blue dot in yellow cluster). Correctly clustered trials correspond to trials with typical spike patterns for a given column. Incorrectly clustered trials are then trials with abnormal spike patterns, that more closely resemble the spike patterns on trials of a different column.  

We hypothesized that the paths taken by the animal on incorrectly clustered trials would more closely resemble the paths in the predicted cluster as opposed to the true cluster (i.e., if a blue trial is grouped into the yellow cluster, we expect the path on that trial to look more like a yellow path ``predicted cluster" than a blue path ``true cluster" - see Supplement Figs.~S3a,b). We used the Fréchet distance to measure the distance between paths. As a baseline, we computed the distance between paths on trials correctly assigned to the same true cluster (see Supplement Fig.~S3c ``correct to true cluster"). This provides a value for the distance between the paths on all correct yellow column trials that fall into the yellow column cluster, for example. For each incorrectly clustered trial, we then computed the mean similarity to other paths in its predicted cluster, compared to other paths in its true cluster. As expected, paths on incorrectly clustered trials were significantly more similar to the paths on trials in the predicted cluster than to other trials of the same condition (true cluster) (p$<$0.003, see Supplement Fig. S3). 

Together, this shows that transient spike patterns in the navigation epoch of the task closely reflect the navigation strategy taken by the subject on individual trials -- so closely, in fact, that the observed spike pattern can predict the specific type of error made on individual trials. 

\subsection{Mathematical Details}
\label{sec:mathematical_details}

\subsubsection{Definitions}

\textbf{Spike pattern:} $ s \subset S^1 \times \{1,...,N\}$ is the collection of points $s:=\{ (x,y)_j \}_{j=1}^M$ that represent $M$ spikes fired by a population of $N$ neurons during some window of time, of length $T$ seconds. The $j^{th}$ element, $(x,y)_j$ corresponds to the spike phasor $x_j \in S^1$ fired by neuron $y_j \in \{1,...,N\}$ at time $t_j$ within the window of analysis. The recorded spike time, $t_j$, is transformed into a spike phasor following $x_j := e^{2\pi \i \frac{t_j}{T} - \pi}$. Let $\mathcal{S}:=\{s\} \subseteq \mathcal{P}(S^1 \times \{1,...,N\})$ denote the set of all such spike patterns. 

\textbf{Vector spike pattern:} Let $s \in \mathcal{S}$ be a spike pattern of the form $s := \{(x,y)_j\}_{j=1}^N$, where each neuron fires precisely one spike. In this case, we can order the spikes by neuron ID so that $s = \{(x,j)_j\}_{j=1}^N$. With a useful abuse of notation, we will represent these spike patterns as vectors $\vec{s}$ whose $j^{th}$ component is the spike phasor of the $j^{th}$ neuron: $\vec{s} := [x_1,x_2,...x_N]$.

\textbf{Helix:} Let \(\omega^k=e^{-2\pi \i k/N}\) denote the $k^{th}$ $N$-root of unity. For $k = 1,\dots,N$, the $k^{th}$ \textbf{helix} is the vector spike pattern $\vec{h_k} = \{(\omega^{jk},j)\mid j=1,\dots,N\} = [\omega^k,\omega^{2k},\dots,\omega^{Nk}]^T$. 

\subsubsection{Properties of $\Psi$}

Our algebraic approach to spike patterns can be understood as operators applied on the data. With this in mind, these operators have specific properties that are necessary to guarantee that our approach can be successfully used on spiking data.  

\textbf{(1)} : $ \Psi_H(\vec{h_k}) = \vec{e}_k $. In particular, $ \| \Psi_H(\vec{h_k}) \| = 1 $ and Arg$( \Psi_H(\vec{h_k}) ) = \vec{0}$. 
\\
\\
The $n^{\text{th}}$ element of $\Psi_H(\vec{h_k})$ is given by: 
\begin{align}
\nonumber
    \mu_n &= \frac{1}{N} \sum_{j=1}^M x_j \cdot (\vec{h}_n)_{y_j}^* \text{ from (Eq. \eqref{eq:mu_k}) }, \\ \nonumber
    &= \frac{1}{N} \sum_{j=1}^N x_j  \cdot (\vec{h}_n)_{y_j}^* \text{ since $M=N$ for helices }, \\ \nonumber
    &= \frac{1}{N} \sum_{j=1}^N \omega^{jk} \cdot (\vec{h}_n)_{j}^* \text{ by the definition of $h_k$},\\ 
    &= \frac{1}{N} \sum_{j=1}^N \omega^{jk} \cdot (\omega^{jn})^* \text{ by the definition of $h_n$}.
\end{align}
 When $n=k$, we have that $\mu_n = \frac{1}{N} \sum_{j=1}^N \omega^{jk} (\omega^{jk})^* = \frac{1}{N} \sum_{j=1}^N 1 = 1$. When $n \neq k$, the contribution sums to $0$ by symmetry in the complex plane.  
\\
\\
\textbf{(2)} : $ \Psi_H( \vec{h_l} \cup \vec{h_k}) = \vec{e}_l + \vec{e}_k  $
\\
\\
The spike pattern $\vec{h_l} \cup \vec{h_k}$ is the union of $\vec{h_l}$ and $\vec{h_k}$; i.e., the spike pattern in which each cell $j$ spikes twice, at phasors $\omega^{jl}$ and $\omega^{jk}$. The $n^{th}$ element of $\Psi_H( \vec{h_l} \cup \vec{h_k} )$ is given by: 
\begin{align}
\nonumber
    \mu_n 
    &= \frac{1}{N} \sum_{j=1}^{2N}x_j \cdot (\vec{h}_n)_{y_j}^*, \\
    &= \frac{1}{N}( \sum_{j=1}^N \omega^{jl} \cdot (\omega^{jn})^* + \sum_{j=1}^N \omega^{jk} \cdot (\omega^{jn})^*),
\end{align}
by summing first over the spikes in $\vec{h_l}$ and then over the spikes in $\vec{h_k}$. As above, when $n=l$ and when $n=k$, $\mu_n = 1$. For all other $n$, $\mu_n = 0$. 
\\
\\
\textbf{(3)} Let $s_\sigma$ denote a spike pattern obtained from $s$ by rotating each spike phasor $x_j$ by phase $\sigma$. Then $ \Psi_H( s_\sigma) = \sigma \vec{e}_k $. In particular, Arg$( \Psi_H( s_\sigma) ) = \sigma$ in the $k^{th}$ component and $0$ everywhere else. When $\sigma = \pi$, we denote $-s := s_\sigma$. 
\\
\\
\textbf{(4)} $\Psi_H(s)$ is unique up to spike patterns with a very specific geometric relationship. More precisely, two non-empty spike patterns $s$ and $t$ have the same ms-DHT if and only if, for each $j$, the spike phasors of neuron $j$ in $s$ together with the spike phasors for neuron $j$ in $-t$ form an origin-centered polygon. 

For an integer $M\geq 2$, we define an origin-centered polygon ($M$-gon) to be a spike pattern made of precisely $M$ spikes of the same neuron, such that the barycenter of the polygon (or segment, if $M=2$ determined by the spike phasors is the origin: $s=\{(x,n)_j\}_{j=1}^M$ with $\sum_j x_j = 0$.

 The matrix of helices
    \[
    H = \left[\begin{array}{c|c|c|c}&&&\\ \vec{h_1} & \vec{h_2} & \dots & \vec{h_N}\\&&&\end{array}\right]
    \]
    is a Vandermonde matrix of roots of unity with a cyclic shift of the rows and the columns (so that the ones are concentrated in the last row and the last column). Therefore, $H$ is invertible, with inverse $H^{-1} = \frac{1}{N}H^\ast$.

    If we consider a vector spike pattern $\vec{v}$, then $\Psi_H(\vec{v}) = \frac{1}{N}H^\ast \vec{v} = H^{-1}\vec{v}$. Equivalently, $\vec{v} = H\Psi_H(\vec{v})$, hence $\vec{v}$ can be uniquely and unambiguously reconstructed from its helix transform $\Psi_H(\vec{v})$.
    
    For an arbitrary spike pattern $s$, consider the vector $\widetilde{s}\in\C^N$ whose $j$-th component is the sum of all the phasors of the spikes of the $j$-th neuron in $s$. In general, $\widetilde{s}$ is not a vector spike pattern, because its components do not have modulus $1$; however, we can still (formally) apply the helix transform to $\widetilde{s}$, and $\Psi_H(s) = \Psi_H(\widetilde{s})$. Since this formal helix transform $\Psi_H$ is an isomorphism of vector spaces, $\widetilde{s}$ can be uniquely and unambiguously reconstructed from the helix transform $\Psi_H(\widetilde{s})=\Psi_H(s)$. Moreover, for two spike patterns $s$ and $t$,
    \begin{align}
    \nonumber
    \Psi_H(s) = \Psi_H(t) &\Leftrightarrow \Psi_H(\widetilde{s}) = \Psi_H(\widetilde{t}) \\ \nonumber
     & \Leftrightarrow \widetilde{s} = \widetilde{t} \\
     \nonumber
     & \text{ (as $\Psi_H$ is an isomorphism $\C^N\to\C^N$)}\\
    & \Leftrightarrow \widetilde{s} - \widetilde{t} = 0 \text{ as vectors in }\C^N
    \end{align}
    which happens if and only if, for each $j$, the spike phasors of neuron $j$ in $s$ together with the spike phasors of neuron $j$ in $-t$ form an origin-centered polygon.

The meaning of the theorem from the point of view of reconstructing a spike pattern from its helix transform is that the reconstruction is possible up to origin-centered polygons in each neuron-coordinate.
\\
\\
\textbf{(5)} Distances between spike patterns determined by the ms-DHT are preserved under cell relabeling. 

In more detail, let $s$ and $t$ be spike patterns. Consider the corresponding spike patterns $\Tilde{s}$ and $\Tilde{t}$ obtained by permuting the neuron IDs, so that each neuron is assigned to a new number. For a neuron with ID $y$ in $s$ and $t$, let $\textbf{p}(y)$ denote the new ID of that neuron in $\Tilde{s}$ and $\Tilde{t}$. 

Consider the $k^{th}$ element of $\Psi_H(s)$: \\$\mu_k = \frac{1}{N} \sum_{j=1}^M x_j (\vec{h_k})^*_{y_j} = \frac{1}{N} \sum_{j=1}^M x_j (\omega^{k y_j})^*$. That same element of  $\Psi_H(\Tilde{s})$ is then $\frac{1}{N} \sum_{j=1}^M x_j (\omega^{k \cdot \textbf{p}(y_j)})^*$. Cell relabeling applies a specific rotation to each term in the sum, and thus modifies each component of the vector $\vec{\mu}$.  

While $\Psi_H(s) \neq \Psi_H(\Tilde{s} )$, we have that $||\Psi_H(s)|| = ||\Psi_H(\Tilde{s})||$, so $|| \Psi_H(s) - \Psi_H(t) || = || \Psi_H(\Tilde{s}) - \Psi_H(\Tilde{t} )||$. See Supplement, Fig. S4 for details. 

\section*{Data and Code Availability}
Data and code used in this study will be made available at https://github.com/mullerlab/buschEAmathematical upon publication. 

\section*{Acknowledgments}
This work was supported by BrainsCAN at Western University through the Canada First Research Excellence Fund (CFREF), NIH U01-NS131914, NIH R01-EY028723, the NSF through a NeuroNex award (\#2015276), the Natural Sciences and Engineering Research Council of Canada (NSERC), Compute Ontario (computeontario.ca), Digital Research Alliance of Canada (alliancecan.ca), and the Western Academy for Advanced Research. A.B. gratefully acknowledges support from the NSERC PGS-D and BrainsCAN Graduate Fellowship Program. J.A.M. was supported by a Banting Postdoctoral Fellowship, a Vector Institute Postgraduate Affiliation, and a BrainsCAN Postdoctoral Fellowship (CFREF). J.A.P. was supported by the Canada Research Chair program.

%

\end{document}